\documentclass[pdflatex,sn-mathphys-num]{sn-jnl}

\usepackage{bm}
\usepackage{comment}
\usepackage{tabularx}
\usepackage{tabulary}
\usepackage[usenames,dvipsnames]{color}
\usepackage[normalem]{ulem}

\usepackage{graphicx}
\usepackage{multirow}
\usepackage{amsmath,amssymb,amsfonts}
\usepackage{amsthm}
\usepackage{mathrsfs}
\usepackage[title]{appendix}
\usepackage{xcolor}
\usepackage{textcomp}
\usepackage{manyfoot}
\usepackage{booktabs}
\usepackage{algorithm}
\usepackage{algorithmicx}
\usepackage{algpseudocode}
\usepackage{listings}
\usepackage{caption}

\theoremstyle{thmstyleone}

\theoremstyle{thmstyletwo}

\theoremstyle{thmstylethree}

\usepackage{lineno}
\begin{document}
\title[Observation of PISN mass gap]{Gravitational-wave constraints on the pair-instability mass gap and nuclear burning in massive stars}
\author*[1]{Fabio Antonini}\email{antoninif@cardiff.ac.uk}
\author[1,2]{Isobel M. Romero-Shaw} 
\author[3]{Thomas Callister}
\author[1]{Fani Dosopoulou}
\author[4]{Debatri Chattopadhyay}
\author[5,6]{Yonadav Barry Ginat}
\author[7,8]{Mark Gieles}
\author[9,10]{Michela Mapelli}

\affil[1]{Gravity Exploration Institute, School of Physics and Astronomy, Cardiff University, Cardiff, CF24 3AA, UK}
\affil[2]{H. H. Wills Physics Laboratory, Tyndall Avenue, Bristol BS8 1TL, UK}
\affil[3]{Kavli Institute for Cosmological Physics, The University of Chicago, Chicago, IL 60637, USA}
\affil[4]{Center for Interdisciplinary Exploration and Research in Astrophysics (CIERA) and Department of Physics \& Astronomy, Northwestern University, 1800 Sherman Ave, Evanston, IL 60201, USA}
\affil[5]{
Rudolf Peierls Centre for Theoretical Physics, University of Oxford, Parks Road, Oxford, OX1 3PU, United Kingdom
}
\affil[6]{New College, Holywell Street, Oxford, OX1 3BN, United Kingdom
}
\affil[7]{
ICREA, Pg. Lluís Companys 23, E08010 Barcelona, Spain;
}
\affil[8]{
Institut de Ciències del Cosmos (ICCUB), Universitat de Barcelona (IEEC-UB), Martí Franquès 1, E08028 Barcelona, Spain
}
\affil[9]{Universit\"at Heidelberg, Zentrum f\"ur Astronomie (ZAH), Institut f\"ur Theoretische Astrophysik, Albert-Ueberle-Str. 2, 69120, Heidelberg, Germany
}
\affil[10]{
Physics and Astronomy Department Galileo Galilei, University of Padova, Vicolo dell'Osservatorio 3, I--35122, Padova, Italy}

\abstract{
Pair-instability should prevent the direct formation of black holes above about $50M_\odot$ creating a ``pair-instability'' mass gap. Yet gravitational-wave observations have detected black holes in this mass range. These systems can be explained with uncertainties in massive-star evolution, or hierarchical mergers in stellar clusters, which are expected to produce large spins with isotropic orientations.
Here we present  evidence for the pair-instability mass gap in the LIGO--Virgo--KAGRA fourth transient catalog, with a lower edge at $44.3^{+5.9}_{-3.5}\,M_\odot$. We also obtain a measurement of the ${}^{12}\mathrm{C}(\alpha,\gamma){}^{16}\mathrm{O}$ reaction rate, yielding an $S$-factor of $268^{+195}_{-116}\,\mathrm{keV\,b}$, a parameter critical for modeling helium burning and stellar evolution.  
The  data reveal two populations: a low-spin group with no black holes above the gap, and a high-spin, isotropic group that extends across the full mass range and occupies the gap, consistent with hierarchical mergers. 
These findings 
are consistent with pair-instability playing a role in shaping the black hole mass spectrum, point to a  connection between gravitational wave astronomy and nuclear astrophysics, and highlight dense stellar clusters as key environments in the growth of black holes. 
}

\maketitle

Gravitational-wave observations of binary black holes have opened a new window onto massive-star evolution \cite{2019PhRvX...9c1040A, 2019ApJ...882L..24A, Abbott:2020gyp, 2021arXiv211103606T, 2021PhRvX..11b1053A}, but population inferences remain hampered by uncertainties in binary physics and initial conditions \cite[e.g.,][]{Spera2015a, belczynski2016, Stevenson2015, Marchant2016}. A central issue is whether (pulsational) pair-instability supernovae (PISN) carve out a gap in the black hole birth mass distribution \cite[e.g.,][]{Woosley2021,2021MNRAS.502L..40F,2020ApJ...890..113B}; theory predicts pulsations for He cores $\sim40$–$65\,M_\odot$ and full disruption above $\sim65\,M_\odot$, suppressing black hole formation in the $\sim40$–$130\,M_\odot$ range \cite{Woosley2016, 2017MNRAS.470.4739S, 2020ApJ...902L..36F,2023MNRAS.526.4130H,2019ApJ...887...53F,2019ApJ...887...72L}. 
Gravitational-wave observations have so far revealed no sharp deficit {of black holes in this  mass range (the so-called PISN mass gap)} \cite{2019ApJ...882L..24A,Abbott:2020gyp,2021ApJ...913L..23E,LVKCollab2023,ray_nonparametric_2023,2024PhRvX..14b1005C,2025PhRvD.112b3531A}, motivating scenarios that populate the gap \cite[e.g.][]{2021MNRAS.502L..40F,2020ApJ...902L..36F,2022MNRAS.516.2252O,2024MNRAS.529.2980W}, or raising the possibility that the gap may not exist at all \cite[e.g.][]{2021MNRAS.501.4514C}.

Dynamical environments (e.g., globular and nuclear clusters or active galactic nuclei disks) can produce merger remnants that merge again, yielding higher spins \cite{PhysRevD.78.044002,PhysRevLett.96.111101,PhysRevLett.96.111102, 2023MNRAS.526.4908C} with isotropic orientations \cite{2006ApJ...637..937O,2016ApJ...831..187A,Rodriguez2015a}, and populating the PISN mass gap.
For binaries in which the primary component was produced by a previous merger, the effective
combination of the two component spins projected parallel to the orbital angular momentum~\cite{2008PhRvD..78d4021R,PhysRevLett.106.241101}, $\chi_{\rm eff}$ ---the best measured spin parameter from data--- is expected to be broad and  symmetric around zero. 
Ref.~\cite{2025PhRvL.134a1401A} showed that this distribution is  largely independent of model assumptions and uncertainties, and derived an approximately uniform form with \(|\chi_{\rm eff}|\!\lesssim\!0.5\). 
This bound follows from the fact that merger remnants are expected to have a nearly universal dimensionless spin magnitude, $a_{\rm rem}\sim 0.7$, as predicted by general relativity~\cite{PhysRevD.78.044002}. 
Assuming a negligibly spinning companion,  $\chi_{\rm eff}$  satisfies
$|\chi_{\rm eff}| \lesssim \frac{m_{\rm rem} a_{\rm rem}}{m_{\rm rem}+m_2}.$
For $m_2 \simeq 0.5\,m_{\rm rem}$, as expected in dynamical formation~\cite{2019PhRvD.100d3027R}, this yields
$|\chi_{\rm eff}| \lesssim 0.47.$
The presence and location of the PISN lower mass limit \(\tilde{m}\) can therefore be inferred from the primary mass \(m_1\) at which the \(\chi_{\rm eff}\) distribution transitions to this broad, symmetric form.

\begin{figure}[h!]
\includegraphics[height=0.7\columnwidth]{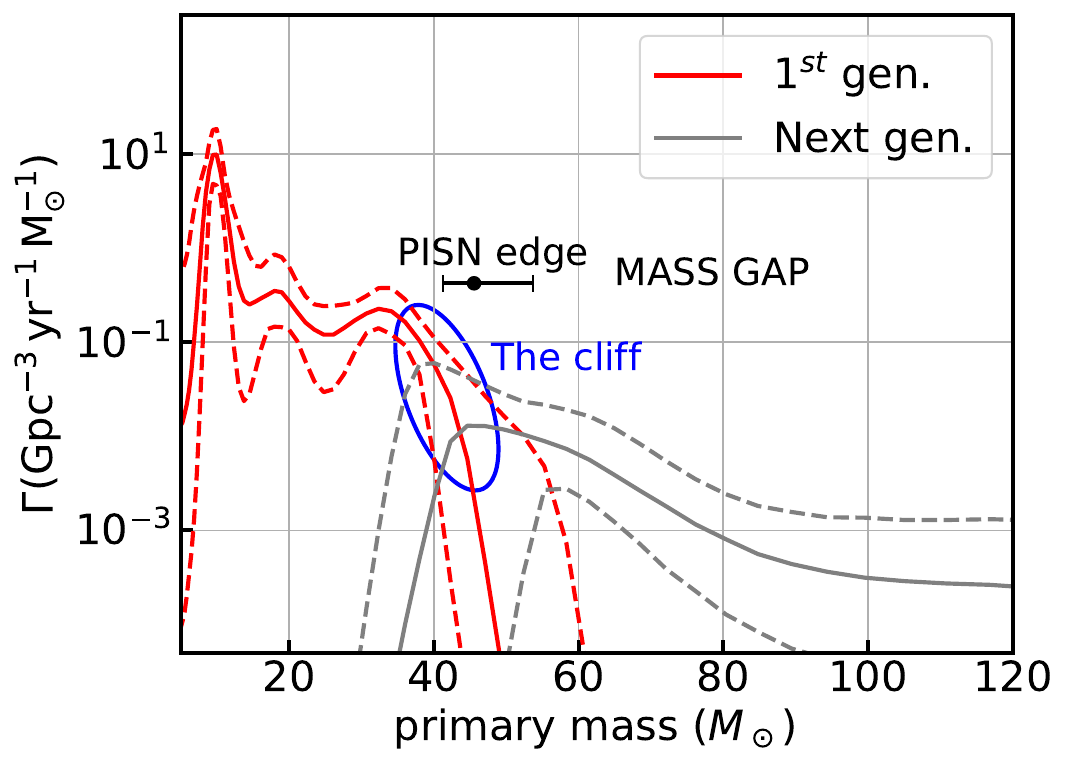} 
    \caption{ {\bf The primary black hole mass spectrum.} ~{Merger rate as a function of primary black hole mass in binaries below (red) and above (black) the truncation mass $\tilde{m}$ separating low- and high-spin populations, calculated at a reference redshift $z=0.2$. 
Given the merger-rate density per unit primary mass at the reference redshift,
$
\Gamma(m_1)
$,
the  primary-mass distribution for the two populations below and above  $\tilde m$ are  reconstructed from the posterior samples as
$\Gamma_{<\tilde m}(m_1)
=
\bigl[1-\eta(m_1)\bigr]\Gamma(m_1)$
and
$\Gamma_{>\tilde m}(m_1)
=
\eta(m_1)\Gamma(m_1)$, respectively.
    Here, $\eta(m_1)$ is a sigmoid mixing function that sets the relative
contribution of the two population components as a function of primary mass,
with $\eta(\tilde m)=0.9$. 
    This model yields a total merger rate at this redshift of $33.4^{+13.3}_{-8.4}\,{\rm Gpc^{-3}\,yr^{-1}}$, consistent with \cite{LVKpop_inprep}.    
    The inferred $\tilde{m}$ is marked by the black point. Solid lines indicate the median merger rate and dashed lines the $10^{\rm th}$–90$^{\rm th}$ percentiles. 
    A mass gap in the low-spin population, isotropic spins above $\tilde{m}$, and a sharp drop in the merger-rate  at the same mass value (the cliff) are all features consistent with a PISN gap populated by hierarchical mergers in dense star clusters. 
    }
    }
    \label{fig:fig1}
\end{figure}

We perform hierarchical Gaussian-process population inference on the fourth LIGO--Virgo--KAGRA gravitational wave transient catalog  \cite[GWTC-4;][]{LVKcat_inprep,LVK_GWTC4_2025} to map black hole spin as a function of primary mass. 
With the source catalog now more than twice as large, we obtain tight constraints and are able to probe new features of the population. 
We fit the $\chi_{\rm eff}$ distribution to a mixture model comprising a Gaussian distribution, representing the bulk of the
population at $m_1 \lesssim \tilde{m}$, and a higher mass spin distribution  described via a non-parametric Gaussian process prior.

We identify a transition at $\tilde{m}=45.6^{+12.7}_{-5.6}\,M_\odot$ (90\% confidence), separating the two  populations. The differential merger rate as a function of primary mass for the two populations are given in Fig.~\ref{fig:fig1}, while
the inferred spin distributions  are given in Supplementary Information.
Below $\tilde{m}$, the data are well described by a single narrow Gaussian $\chi_{\rm eff}$ distribution, $\log_{10} \sigma=-1.15^{+0.13}_{-0.15}$, with small and positive mean, $\mu=0.04^{+0.02}_{-0.02}$, consistent with first-generation black holes.
The merger rate of this population drops to zero above $\tilde{m}$, implying a gap in the mass spectrum {of first generation mergers}. In contrast, the population above $\tilde{m}$ exhibits a much broader spin distribution with a median value 
$\langle \chi_{\mathrm{eff}}\rangle=
0.11^{+0.197}_{-0.243}$. Under the proposed model, the data are consistent with 
a $\chi_{\rm eff}$ distribution that is symmetric about zero and  with the expected distribution of second-generation mergers formed dynamically in dense stellar environments.

The precise measurement of  $\tilde{m}$ means that the mixture model is strongly favored over models in which black holes
of all masses share the same spin distribution. 
To quantify support for the transition-mass model, we compute the Bayes factor relative to the  default model used by the LIGO-Virgo-KAGRA collaboration \cite[e.g.,][]{LVKpop_inprep} in which the full  $\chi_{\rm eff}$ distribution is described by a single truncated Normal, $p(\chi_{\rm eff})= \mathcal{N}(\chi_{\mathrm{eff}};\, \mu, \sigma)$, with no distinct high-mass component. We sample this hierarchical model using the same event set, likelihood, and inference framework as in the main analysis.
We obtain a Bayes factor $B > 10^4$ in favor of the model with a separate high-mass spin component.




\begin{figure*}
    \centering  
\includegraphics[width=1.\textwidth]{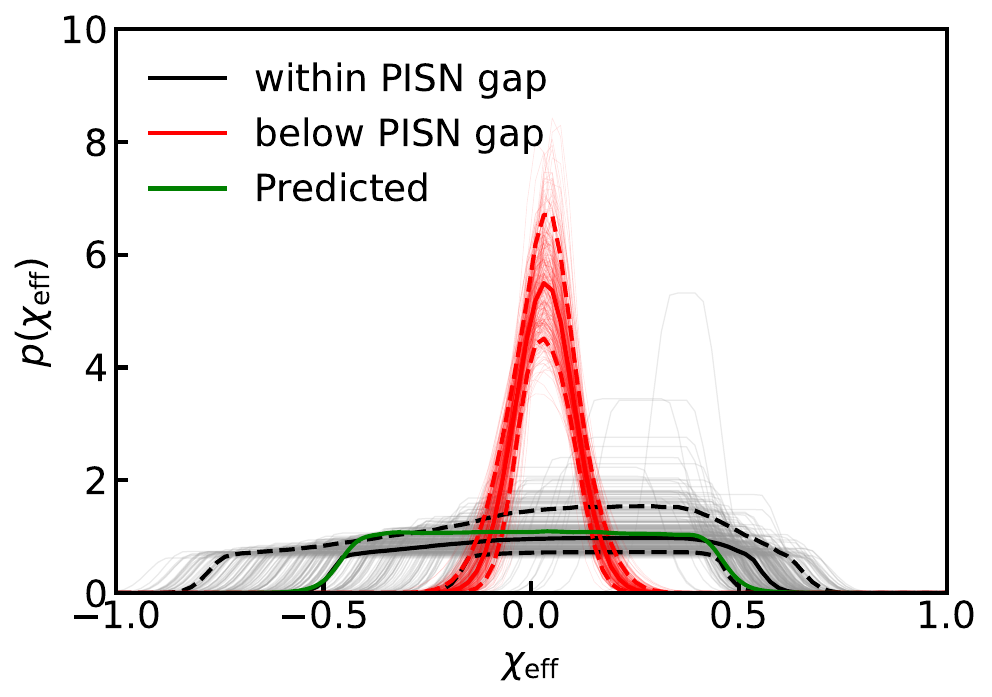}
    \caption{
    {\bf 
The $\chi_{\rm eff}$ distribution   below and within the PISN mass gap.} Thick black lines show the $\chi_{\rm eff}$ distribution for the high-mass population, $m>\tilde{m}$, modeled as a uniform distribution with independent bounds, while thick red lines correspond to the $\chi_{\rm eff}$ distribution  of the low-mass population, $m<\tilde{m}$.
Solid  lines are median, while dashed lines show 10\% and 90\% of the distributions.
 The  distribution for $\chi_{\rm eff}$ {in} the PISN mass gap predicted under a hierarchical formation scenario is shown in green \cite{2023MNRAS.522..466A}. 
 Light color traces correspond
to a single draw from the posterior distribution, providing a visual representation of the sample support from which the confidence intervals are constructed.
}
    \label{par}
\end{figure*}

\begin{figure}
\includegraphics[height=.56\columnwidth]{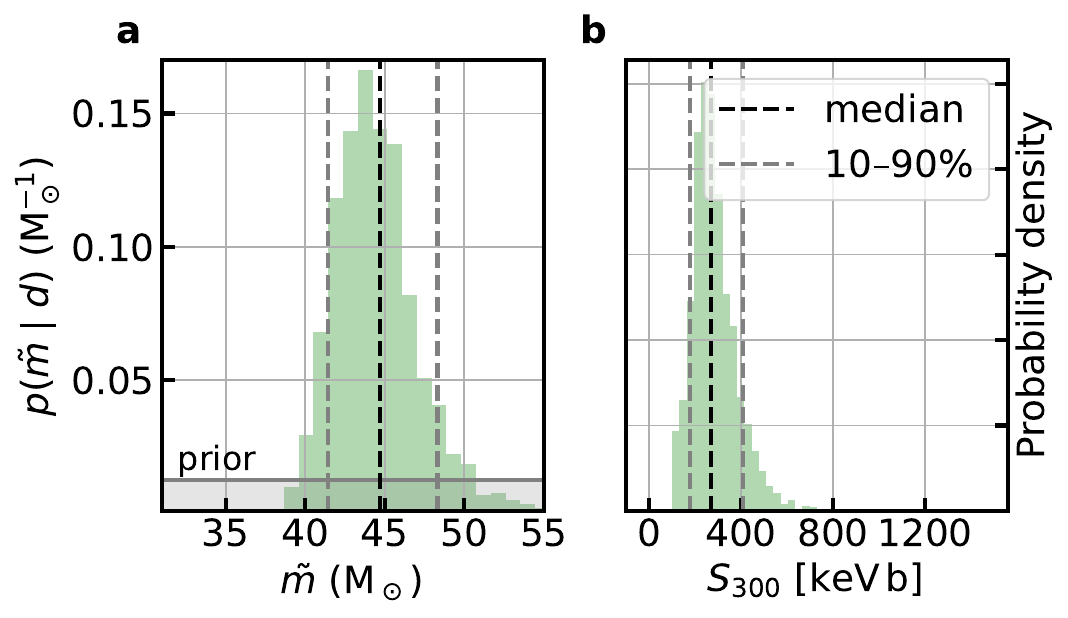} 
    \caption{
 \textbf{Constraints on the PISN transition mass and the $^{12}\mathrm{C}(\alpha,\gamma)^{16}\mathrm{O}$ reaction rate.} Posterior distribution
of the primary mass value separating the two black hole populations, $\tilde{m}$. ~{\bf b}, The corresponding posterior of the astrophysical factor, $S_{300}$. The latter is derived from  $\tilde{m}$ using this as the value of the lower edge of the PISN
mass gap.
    }
    \label{fig:fig2}
\end{figure}

\begin{figure}
\includegraphics[height=1\columnwidth]{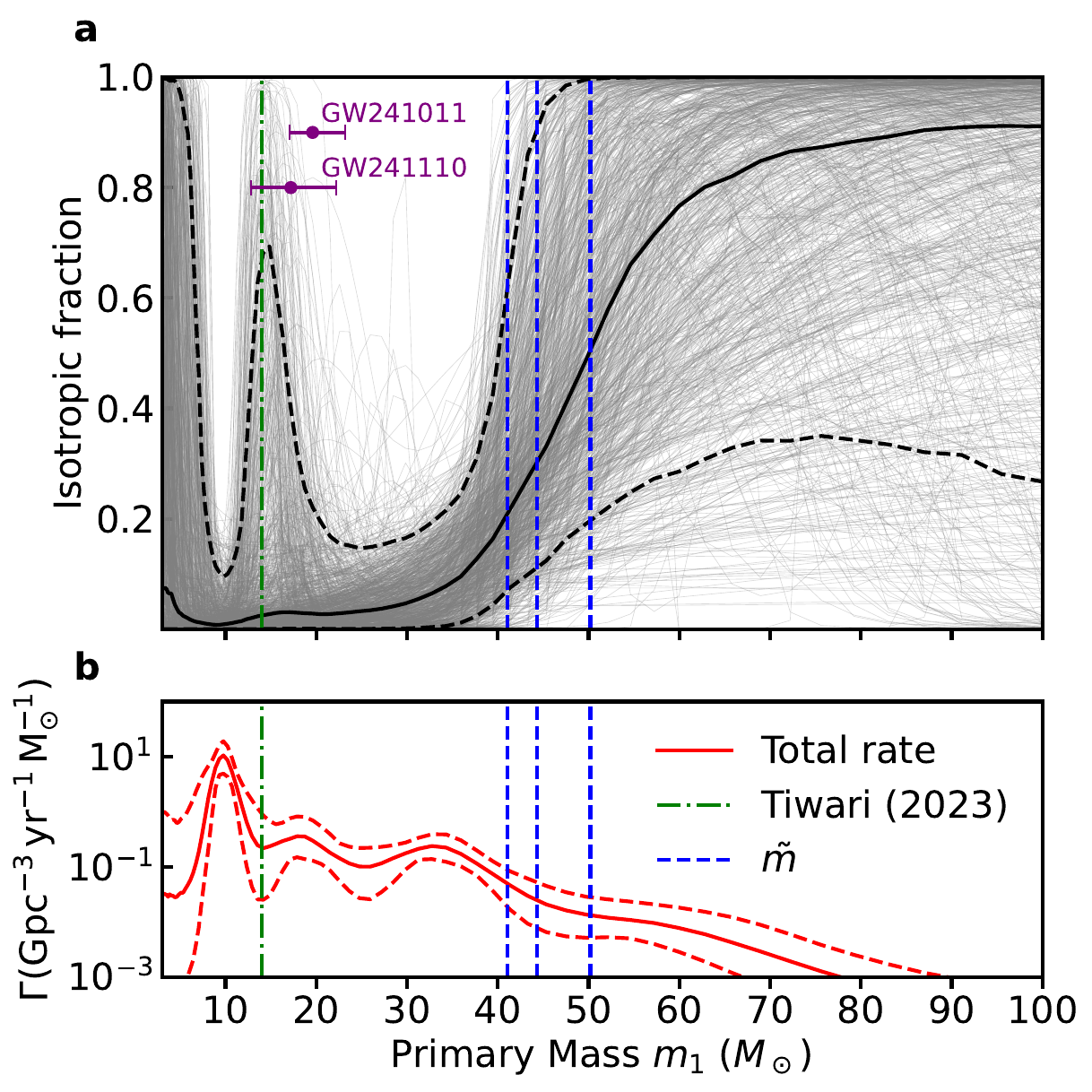} 
    \caption{
 \textbf{Mass-dependent mixture fraction between the two black hole populations.} {\bf a}, The mixture fraction between the two black hole populations, modelled non-parametrically as a function of mass.  Median, 10\% and 90\% percentiles are shown. Light black lines are individual traces. {\bf b}, For reference, the total differential merger rate as a function of primary black hole mass is shown.
A transition above $\simeq 50 M_\odot$ is clearly recovered by this  analysis. We also find a possible indication of an isotropic population at a primary mass of $\sim 14\,M_\odot$, although it is not statistically required by the data. { This aligns with the mass dip in \cite{2024MNRAS.527..298T} and with GW241011 and GW241110, two O4b events interpreted as hierarchical mergers \cite{2025ApJ...993L..21A}. Together, these support our interpretation of a low-mass ``valley'' which is populated by hierarchical mergers.} Below $\sim 10M_\odot$, the posterior broadens, reverting to the prior due to the absence of sources there.
}
    \label{fig:fig3}
\end{figure}




The overall mass distribution shows several features. There are peaks at  $\simeq 10\,M_\odot$, $18\,M_\odot$, and $38\,M_\odot$, as also reported by \cite{LVKpop_inprep} ---these peaks were previously identified by Ref.~\cite{2021ApJ...913L..19T}. There is a drop of nearly two  orders of magnitude in the total merger rate at  $\sim 40\,M_\odot$
(we name this feature ``the cliff"  in Figure \ref{fig:fig1}).
A rapid decline in the merger rate followed by a plateau (or a shallower decline) at 
$\tilde{m}$, as we found, is a generic feature of cluster formation models \cite{2023MNRAS.522..466A,2025PhRvL.134a1401A}. The plateau starts at the onset of pair-instability supernovae and it is due to the emergence of binaries with components formed from previous mergers.  
{  According to these models,  the edge of the transition between the two populations can  be estimated from the primary mass distribution alone as the value of maximum curvature above $30M_\odot$ in the differential merger rate, $\Gamma(m_1)$  
(See Methods\ \ref{PISNclusters}). We measure this transition from the data
at $\simeq 42M_\odot$, consistent with the independent  estimate based on the spin transition.
A comprehensive interpretation of the cliff may require accounting for multiple concurring physical processes, as discussed in Supplementary Information.}

 The non-parametric analysis over a large dataset gives us  confidence that a transition to a broader and more uniform distribution exists in the population. Motivated by this, we introduce a more informed parametric model in which the high-mass 
population is described by a uniform distribution with independent bounds, and use
this model in what follows (unless otherwise specified).
As before, the distribution
below $\tilde{m}$ is well described by a Normal distribution with
positive mean, $\mu=0.05^{+0.02}_{-0.02}$, and small dispersion,
$\log_{10} \sigma=-1.29^{+0.18}_{-0.17}$. The distribution
above $\tilde{m}$ is a broad distribution for which we 
 infer upper and lower bounds of $\chi_{\rm eff,\max} = 0.49^{+0.14}_{-0.12}$ and $\chi_{\rm eff,\min} = -0.41^{+0.40}_{-0.35}$, respectively. {The lower bound is less well constrained than the upper bound, primarily due to selection effects and parameter-estimation uncertainties. Nevertheless, we find that $\chi_{\rm eff,\min} < 0$ with $98.4\%$ credibility, providing strong evidence for misaligned spins in the population. The median value $\langle \chi_{\rm eff} \rangle = 0.02^{+0.18}_{-0.17}$, indicates that the distribution is statistically consistent with being symmetric around zero.} 
Figure~\ref{par} shows the corresponding recovered $\chi_{\rm eff}$ distributions.

From the same parametric model we infer a characteristic mass scale of $\tilde{m} = 44.3^{+5.9}_{-3.5}\,M_\odot$, a value close to the lower edge of the {pair instability} mass gap reported in theoretical and numerical studies~\cite{Woosley2016,2019ApJ...887...72L,2023MNRAS.526.4130H,2019ApJ...887...53F}. The posterior distribution of $\tilde{m}$ is shown in the left panel of Figure~\ref{fig:fig2}.

Our results are consistent with a depletion of first-generation, low-spin black holes above 
$\tilde{m} \simeq 45\,M_\odot$.
We reported a similar transition at $46^{+7}_{-6}\,M_\odot$  in GWTC-3 using 69 sources, with 11 having the $90\%$ of the $m_1$ posterior distribution above $45\,M_\odot$ after population reweighting \cite{2025PhRvL.134a1401A}, 
and  a  transition  to a higher-spin population was identified by others \cite{2022ApJ...941L..39W, 2023arXiv230302973L}.
The consistent recovery in the new, larger catalog containing 153 sources, with $34$ above  $45M_\odot$,  demonstrates that the feature is strongly driven by the data and it is not a statistical fluctuation.  
Constraints on the \(\chi_{\rm eff}\) distribution of the high-mass population are now tighter and more consistent with our theoretical interpretation. Using GWTC-3 we inferred \(\chi_{\rm eff,max}=0.57^{+0.21}_{-0.19}\) (uniform model), and could not decisively exclude \(\chi_{\rm eff,max}=1\) or \(\chi_{\rm eff,max}<0\) \cite{2025arXiv250609154A}. With the expanded data set we can instead  rule out both \(\chi_{\rm eff,max}=1\) and \(\chi_{\rm eff,max}<0\) and infer $\chi_{\rm eff,\max}\simeq 0.5$. This is notable because hierarchical mergers is the only astrophysical pathway that robustly predicts this upper bound~\cite{2025PhRvL.134a1401A}.
The low-\(\chi_{\rm eff}\) tail is also better constrained. In GWTC-3, \(\chi_{\rm eff,min}\) and the left tail were sensitive to prior choices  \cite[e.g., Fig.~5 in][]{2025arXiv250609154A}; with the larger sample we find \(-1<\chi_{\rm eff,min}<0\) with substantially higher confidence, and a peak near the expected value \(\chi_{\rm eff,min}\sim -0.5\).

Together, these robust features indicate that the data are consistent with a hierarchical origin of the high-mass population: a mass gap 
in the low mass/low spin population, 
the onset of an isotropic and highly spinning population above $\tilde{m}$, the sharply defined upper bound of the $\chi_{\rm eff}$ distribution at $\simeq 0.5$, and the steep decline in the total merger rate (the ``cliff'') near the transition. The transition mass itself, $\tilde{m}\simeq 45\,M_\odot$,  matches stellar evolution predictions for the onset of the pair-instability mass gap. 
Recent studies
Recent studies \cite{2025arXiv250602250S,2025arXiv250819208M} have identified a sharp decline in the merger rate---or possibly even a ``gap,'' as suggested by concurrent work \cite{2025arXiv250904151T}---for systems with secondary masses above
$\simeq 45\ M_\odot$, which is expected due to the rarity of binaries in which both components are  second-generation black holes. Although PISN-gap mergers are expected mainly to involve a first-generation BH and a merger remnant, we note that binaries where both BHs are merger remnants should also occur, as has been suggested for GW190521 \cite[e.g.,][]{2021MNRAS.502.2049L}.
Together, these independent lines of evidence  favour hierarchical mergers as the most likely origin of the high-mass population.
Such signatures are difficult to explain through isolated binary evolution, but arise naturally if the high-mass population is built from hierarchical mergers in dense stellar environments.
With this level of confidence, we can now use this result to place direct constraints on massive-star evolution and the physics of the pair-instability process.

The location of the PISN boundary is ultimately set by stellar evolution physics, and in particular by the relative abundances of carbon and oxygen in the cores of very massive stars prior to collapse. These abundances depend  on the \({}^{12}\mathrm{C}(\alpha,\gamma){}^{16}\mathrm{O}\) reaction rate, which governs the conversion of carbon into oxygen during helium burning \cite{Woosley2016,2019ApJ...887...72L,2023MNRAS.526.4130H,2019ApJ...887...53F}.
A higher rate enhances oxygen production, leading to larger oxygen-rich cores and, consequently, to PISN occurring at lower stellar masses. Conversely, a lower rate leaves behind more carbon, shifting the onset of pair instability to higher progenitor masses. Thus,  measurements of the PISN mass gap from gravitational-wave observations of black hole mergers can provide an astrophysical constraint on the \({}^{12}\mathrm{C}(\alpha,\gamma){}^{16}\mathrm{O}\) cross section, a quantity that remains one of the most important nuclear-physics uncertainties in massive stellar modeling \cite{2013ApJS..207...18S,2020ApJ...902L..36F}.

We assume that $\tilde{m}$
is the lower edge of the PISN mass gap, and
follow \cite{2024ApJ...976..121G,2020ApJ...902L..36F} 
to translate our inferred $\tilde{m}$ posterior into 
an estimate of the corresponding astrophysical $S$-factor at 300~keV, $S_{300}$.
The astrophysical $S$-factor rewrites a nuclear reaction cross section by factoring out the strong Coulomb-barrier dependence, 
$
\sigma(E)=\frac{S(E)}{E}\,e^{-2\pi\eta},
$
where $E$ is the center-of-mass energy and $\eta$ the Sommerfeld parameter. 
 We obtain 
$S_{300}=268^{+195}_{-116}\,\mathrm{keV\,b}$ (90\% credibility); we plot this probability distribution in Figure \ref{fig:fig2}. 
This estimate is consistent, within uncertainties, with recent nuclear physics determinations \cite{An2015,deBoer2025,Shen2023}.

Our inference of the ${}^{12}\mathrm{C}(\alpha,\gamma){}^{16}\mathrm{O}$ $S$-factor from gravitational-wave data provides a novel, astrophysical constraint on a parameter that has long been central to stellar evolution theory. 
It relies solely on the assumption that the population with mass $\gtrsim 45M_\odot$  consists entirely of second- (or higher-) generation black holes.
Although direct nuclear physics experiments have yielded estimates with large uncertainties \cite{2020ApJ...902L..36F}, our measurement achieves substantially tighter bounds, enabled by the sensitivity of the black hole mass spectrum to the details of helium burning. This improvement has wide-ranging implications: the carbon-to-oxygen ratio set by this reaction influences the core structure of massive stars, and thus affects the predicted rate of core-collapse supernovae, the maximum masses of neutron stars, and the fate of red supergiants. It also governs the composition of white dwarfs \cite[e.g.,][]{1997ApJ...486..413S}, with consequences for Type~Ia supernova explosions \cite[e.g.,][]{2019ApJ...878...49W}, and shapes the nucleosynthetic yields that feed into Galactic chemical evolution \cite[e.g.,][]{1988ApJ...328..653B}. More broadly, the balance between carbon- and oxygen-rich material determines the conditions for planet formation and the likelihood of forming C-rich versus O-rich planetary systems \cite[e.g.,][]{2011ApJ...743L..16O}. Gravitational-wave astronomy therefore not only constrains the physics of compact objects, but also offers a new window into the nuclear processes that regulate stellar evolution and the chemical enrichment of the Universe.

We now use a non-parametric approach to search for  additional isotropically spinning components in the data and to further test our result of a transition in spin properties at $m_1\simeq45\,M_\odot$.  
We model the mixture fraction
between the low-spin and the high spin and isotropic population as a non-parametric function of the primary mass.
The posterior distribution of the mixture fraction is shown in Figure \ref{fig:fig3}, indicating that the 
fraction of isotropically spinning binaries is consistent with a sharp increase  above $\gtrsim 45\,M_\odot$. 

A new feature appears at $m_1 \simeq 14\,M_\odot$, where the  90\% bound of the mixing fraction rises to $\simeq 0.6$. This coincides with a possible dip in the merger rate—previously identified by \cite{2021ApJ...913L..19T,2024MNRAS.527..298T}. We interpret this as marginal evidence for an additional lower-mass gap in first-generation black holes that {is} populated by black holes formed from a previoous merger \cite{2022ApJ...928..155T}. While consistent with current data, this feature is not  statistically required. Applying the same model to GWTC-3 \cite{2021arXiv211103606T} yields  an upper bound of $\simeq 0.16$, showing that this feature only becomes discernible  with the larger GWTC-4 catalog. 

The data indicate that nearly all primary black holes above $45\,M_\odot$ involved in binary mergers possess high, isotropic spins. Explaining this within stellar evolution would require a mechanism that produces mass-dependent black hole spins at the end of massive star lifetimes while also overcoming the pair-instability gap. The latter might be achieved through reduced stellar winds at low metallicity combined with the collapse of the residual hydrogen--rich envelope during a failed supernova \cite{2021MNRAS.501.4514C,Vink2001}, but no explanation currently exists for the former. Another possibility is that stellar evolution could generate rapidly rotating black holes above $\sim 45\,M_\odot$ through fallback of angular-momentum--rich envelopes.

{Our preferred} explanation is that primary black holes above \(45\,M_\odot\) are the products of repeated mergers in globular clusters \cite{2023MNRAS.522..466A}. The inferred merger rate above this mass therefore provides a strong constraint on the initial cluster density: if all mergers above \(45\,M_\odot\) have this origin, the models of \cite{2023MNRAS.522..466A} imply formation densities of \(\gtrsim 10^4\,M_\odot\,{\rm pc}^{-3}\). Repeated mergers may also occur in active galactic nucleus disks \cite{Bartos2016} or nuclear star clusters \cite{2016ApJ...831..187A}. Because these environments have higher escape velocities than globular clusters, the detailed shape of the \(m_1\) distribution could help determine their relative contributions. After our work appeared on arXiv, Ref.~\cite{2025arXiv251018867R} likewise found a zero-symmetric \(\chi_{\rm eff}\) distribution above \(45\,M_\odot\). They showed that the two populations can also be distinguished by their mass-ratio distributions, with the higher-mass population favoring more asymmetric binaries, consistent with hierarchical formation in clusters. They  noted, however, that the mass-ratio distribution of the high-mass population was not reproduced by their particular globular-cluster models. This could either reflect  uncertainties in the cluster models and in the limited parameter space explored,  or additional channels contributing within the PISN gap.

As the catalog of detected binary black holes continues to expand with future observing runs, constraints on the pair-instability mass gap will sharpen, enabling increasingly stringent bounds on the $^{12}\mathrm{C}(\alpha,\gamma)^{16}\mathrm{O}$ cross section. 
In the coming years, gravitational-wave population inference will thus not only elucidate the astrophysical environments where black holes form and merge, but also offer a new avenue to constrain fundamental nuclear reaction rates that underpin the evolution and fate of massive stars.
At the same time, the identification of a population formed in dense star clusters offers a powerful opportunity to probe their initial conditions and evolutionary pathways across cosmic time.


\section*{Methods}

\section{Population models}
We consider the subset of binary black hole mergers in GWTC-4 with false alarm rates below \(1\,{\rm yr}^{-1}\), consistent with Ref.~\cite{LVKpop_inprep}. 
The data we used are public open data by the LIGO--Virgo--KAGRA collaboration~ \cite{LVK_GWTC2.1_data_quality_2022,LVK_GWTC3_2023,LVK_GWTC4_2025}.

For events first published in \href{https://dcc.ligo.org/LIGO-P1800370/public}{GWTC-1}~\cite{2019PhRvX...9c1040A}, we use the “\texttt{Overall\_posterior}” parameter estimation samples.
For events first published in \href{https://dcc.ligo.org/LIGO-P2000223/public}{GWTC-2}~\cite{2021PhRvX..11b1053A} and 
~\cite{2024PhRvD.109b2001A}, we adopt the “\texttt{PrecessingSpinIMRHM}” samples,  for  events in GWTC-3~\cite{2021arXiv211103606T}, we use the “\texttt{C01:Mixed}” samples available at https://zenodo.org/record/5546663
and for events in GWTC-4 we use the
NRSur7dq4 samples if available \cite{2019PhRvR...1c3015V}, or the
“\texttt{Mixed}”  samples otherwise.
We exclude the events which include at least one component with mass \(< 3\,M_\odot\) and are therefore likely to involve a neutron star~\cite{LVKCollab2023,LVKpop_inprep,LVK_GWTC4_2025}.
This results in 153 events. The detections in GWTC-4 were enabled by a variety of detector improvements
\cite{2020PhRvD.102f2003B,2025PhRvD.111f2002C,2025CQGra..42h5016S,2025Natur.643..955N,PhysRevX.13.041021,2024Sci...385.1318J}.
Selection effects are accounted for using the set of successfully recovered binary black hole injections made publicly available by the LIGO–Virgo–KAGRA collaboration, covering their first four observing runs~\cite{LVKCollab2023,injections,LVKpop_inprep}.
{  Thus, our analysis  accounts for selection effects and measurement uncertainties through the hierarchical Bayesian inference framework, which models the population distribution while marginalizing over individual-event
posteriors.
}

We assume the merger rate density factorizes as
\begin{equation}
\begin{aligned}
R(m_1, m_2, \chi_{\rm eff}; z) &= 
    R_{\rm ref}\,\frac{f(m_1)}{f(20\,M_\odot)} \left(\frac{1+z}{1.2}\right)^\kappa 
 p(m_2 | m_1)\, p(\chi_{\rm eff} | m_1),
\end{aligned}
\end{equation}
where $m_2$ is the secondary mass, and \(R_{\rm ref}\) is the rate per unit mass at \(m_1 = 20\,M_\odot\) and \(z=0.2\).
Our main focus is the conditional spin distribution \(p(\chi_\mathrm{eff}|m_1)\), for which we consider a flexible non-parametric model. 

In our analysis we
simultaneously infer the distributions of binary black hole
primary masses {$m_1$}, mass ratios $q$, and redshifts $z$. 
We model the conditional distribution of the secondary mass \(m_2\) as (e.g.,~\cite{2022ApJ...937L..13C}):
\begin{equation}
\label{eq:pm2}
p(m_2 | m_1) \propto m_2^{\beta_q}, \qquad 2\,M_\odot \leq m_2 \leq m_1.
\end{equation}
Meanwhile, we assume that the volumetric merger rate evolves as a power law in $(1+z)$~\cite{Fishbach_2018,Callister_2020}, such that probability distribution of merger redshifts is
\begin{equation}
\label{eq:pz}
p(z) \propto \frac{1}{1+z} \frac{dV_c}{dz} (1+z)^\kappa.
\end{equation}

In all models, the primary mass spectrum is  modeled non-parametrically with a Gaussian process~\cite[GP;~][]{gelman2013bda3}:
$f(m_1) = \exp[\Phi(\ln m_1)], \qquad 
\Phi(x) \sim \mathcal{GP}\big(0,\,k(x,x';a_m,\ell_m)\big),
$
with a squared-exponential kernel. 
Here,
$a_m$ is the amplitude of the GP (controlling vertical variation), and $\ell_m$ is the length scale (controlling
smoothness), which are 
 treated  as free hyperparameters.
The GP is evaluated on a uniform grid in \(\log m_1\) between \(2\)–\(200\,M_\odot\), and interpolated to event samples and injections.

We model the 
$\chi_{\rm eff}$ distribution as a mixture of two components: a truncated Gaussian
between $[-1, 1]$, describing the bulk of the population at $m_1 \lesssim \tilde{m}$, and a flexible non-parametric distribution at $m_1 \gtrsim \tilde{m}$. The parameter $\tilde{m}$ marks the transition between the two regimes:
\begin{equation} \label{Xeff_GP}
    p(\chi_\mathrm{eff}|m_1) = 
        \left[1-\eta(m_1)\right]\,\mathcal{N}(\chi_\mathrm{eff};\mu,\sigma)
        + \eta(m_1)\,{\exp\left[{\Theta(\chi_{\mathrm{eff}})}\right]}/{\int_{-1}^{1} e^{\Theta(\chi_{\rm eff})}\, d\chi_{\rm eff}}
        ,
    \end{equation}
where 
    \begin{equation}\label{eta}
    \eta(m_1)
    = \left[{1 + {1\over 9}\exp\left(-\frac{(m_1 - \tilde{m})}{M_\odot}\right)}\right]^{-1}\ .
    \end{equation}
This choice ensures that $\eta(\tilde m)=0.9$, i.e., at $m_1=\tilde m$ the Gaussian component contributes 10\% of the total, while the non-parametric component dominates above the transition.
The function $\Theta(\chi_{\mathrm{eff}})$ is generated from a GP,
    $
    \Theta(\chi_{\mathrm{eff}}) \sim \mathcal{GP}\left(0,\, k(\chi_{\mathrm{eff}}, \chi_{\mathrm{eff}}'; \ a_\chi, \ell_\chi)\right),
 $
 with zero mean and a squared-exponential covariance kernel. 
We evaluate these GPs on a regular grid of $N_{\rm bin}=100$ points in $\chi_{\mathrm{eff}}$ within the range $-1$ to $+1$, following \cite{2025arXiv250609154A}. 
We verified that our non-parametric model is sufficiently flexible to recover a narrow \(\chi_{\rm eff}\) distribution if that is what the data prefer. In particular, we model the full population with a GP prior, i.e., we repeat the inference using a model in which 
$p(\chi_{\mathrm{eff}}\,) ={\exp\left[{\Theta(\chi_{\mathrm{eff}})}\right]}$. The GP contracts to  an approximately Gaussian, narrow distribution. This demonstrates that the broader, mass-dependent behavior we infer is data-driven rather than imposed by the model.


Similarly, in the main text we consider a model where the mixing fraction between the two populations,
$\zeta$,  is a non-parametric function of $m_1$ \cite{2025arXiv250609154A}.
Here $\zeta(m_1)$ denotes the fraction of binaries with isotropic spin orientations,
such that
$
p(\chi_\mathrm{eff}\,|\,m_1) = (1-\zeta(m_1))\,\mathcal{N}(\chi_\mathrm{eff};\mu,\sigma) 
+ \zeta(m_1)\,\mathcal{U}(\chi_\mathrm{eff};w)
$, and  $\mathcal{U}(\chi_\mathrm{eff};w)$ is a uniform distribution over 
$|\chi_\mathrm{eff}|<w$, with $w$ the width of the distribution that is also recovered from the data. In this model
we set
\begin{equation}
\zeta(m_1) = S[\Psi(\ln m_1)], 
\qquad 
\Psi(x) \sim \mathcal{GP}\big(0,\,k(x,x';a_\zeta,\ell_\zeta)\big),
\end{equation}
where the Sigmoid function
\begin{equation}
S(x) = \frac{1}{1 + e^{-x}}
\end{equation}  
is applied to ensure $0 \leq \zeta(m_1) \leq 1$.

In our analysis we adopt an additional effective spin model that transitions from a Gaussian to a uniform distribution below and above $\tilde m$, respectively. We treat the bounds of the uniform component, $\chi_{\rm eff,min}$ and $\chi_{\rm eff,max}$, as free parameters inferred from the data:  
\begin{equation}
\label{UL_ind}
p(\chi_{\mathrm{eff}}\,|\,m_1) =
\left[1-\eta(m_1)\right]\,\mathcal{N}(\chi_\mathrm{eff};\mu,\sigma)
        + \eta(m_1)\,\mathcal{U}(\chi_{\mathrm{eff}};\, \chi_{\rm eff,min}, \chi_{\rm eff,max}).
\end{equation}
We place uniform priors on these bounds: $\chi_{\rm eff,max}\sim \mathcal{U}(0.05,1)$ and $\chi_{\rm eff,min}\sim \mathcal{U}(-1,\chi_{\rm eff,max})$.
We plot the joint distribution of the primary black hole mass $m_1$ and the
effective inspiral spin parameter $\chi_{\rm eff}$ under this model in the Supplementary Figure~1.


The priors of the hyperparameters of our models are given in the Supplementary Table 1.

\section{ The pair instability mass gap in star cluster models}\label{PISNclusters}
In the presence of a PISN mass gap, a  feature might be expected to appear in the primary mass distribution at its lower boundary due to a drop in the total merger rate \cite[e.g.,][]{2025A&A...694A.186G}. Using cluster models, we show here 
that this feature takes the form of a transition in the primary-mass distribution, where the merger rate shifts from a steep decline to a flat or even rising trend. The mass at which this transition occurs provides an approximate estimate of the lower edge of the gap. We show that a similar transition is present in the observational data, and that the inferred mass scale is consistent with the value obtained from the spin transition.

For the gravitational-wave rate $\mathcal{R}(m_1)$, we compute the second derivative of the logarithm as
\begin{equation}
	k(m_1) \equiv \frac{\mathrm{d}^2\log \mathcal{R}}{\mathrm{d}(\log m_1)^2}\,, 
\end{equation}
and the curvature 
\begin{equation}
	\kappa(m_1) \equiv \frac{\mathrm{d}^2 \mathcal{R}}{\mathrm{d}m_1^2}\left[ \max \mathcal{R}^2 + \left(\frac{\mathrm{d}\mathcal{R}}{\mathrm{d}m_1}\right)^2\right]^{-3/2}\,.
\end{equation}
While $k$ is dimensionless, $\kappa$ corresponds to the standard definition of the curvature of a curve, and has dimensions of [Mass]$^{-2}$. 
If there is a mass $m_*$ where the rate ceases to be dominated by one population, e.g. first-generation mergers in clusters, and starts to become dominated by another, e.g. second-generation mergers in clusters, and if each population corresponds to a rate which is approximately a power-law, it is natural to expect the derivative of $\mathcal{R}$ to change abruptly at the transition between the two populations. Thus, 
the maxima of $k$ or $\kappa$ would be around $m_*$. 

To investigate this, we consider the  suite of cluster population-synthesis simulations of
Ref.~\citep{2023MNRAS.522..466A} that were obtained
from  the cluster population code \verb"cBHBd", and which  explored a range of model parameters.
These models have been shown to provide a good match to the observed primary mass distribution and rate inferred from the GW data  above $\simeq 30\,M_\odot$.
For each simulation we computed $k$ and $\kappa$, varying the initial cluster half-mass density, the black hole spin distribution, and the inclusion or omission of tidal interaction with an external galactic field. The “zero-spin’’ models were run using the  \verb"cBHBd" settings used in Ref.~\citep{2023MNRAS.522..466A}; the “no-tides’’ models disabled tidal evolution entirely; and in the “beta-function spin’’ models the initial spins were drawn from a $\mathrm{Beta}(2,0.5)$ distribution. 
For each of these three models we explored a grid of 25 initial cluster half-mass densities sampled from  
$10^{4}$ to $10^{6.5}\,M_\odot\,\mathrm{pc}^{-3}$.
For each density, we ran 27{,}216 cluster models—spanning cluster masses from 
$10^{2}$ to $10^{7}\,M_\odot$, 25 metallicities between $1.26\times 10^{-4}$ and $0.025$, 
and a range of formation redshifts.
To mitigate statistical fluctuations, we ran each cluster model with 10 distinct initial random seeds and averaged the results.
Finally, the primary-mass distribution is reconstructed by forward-modelling the full cluster population as in Ref.~\cite{2023MNRAS.522..466A}: we sample dynamical cluster models within a cosmological framework that includes cluster formation histories, metallicity evolution, and the cluster mass function, and use these to generate predictions for the detected astrophysical population.
We smoothed the resulting mass distribution using a Gaussian kernel with width $\sigma_m = 2M_\odot$, though the results are insensitive to the precise choice of smoothing.

For the \verb"cBHBd" 
 simulations, the maximum first-generation primary mass is $m_* = 50 \,M_{\odot}$. The Supplementary Figure 2 displays a histogram of the value of $m_1$ where these measures of the curvature attain their maxima. The maxima are obtained very close to the cutoff mass $m_* = 50 \,M_{\odot}$, with remarkably small variance, irrespective of the model used; this leads us to conclude that this is a robust feature of the cluster channel. 
For $k(m_1)$, we find that the maximum overestimates the value of $m_* = 50\, M_\odot$ expected if $\rho_{h0} > 3\times 10^5\,M_{\odot} \textrm{ pc}^{-3}$; this is because of low-mass second-generation primaries, whose progenitors were below the first-generation peak, become sufficiently prominent at high cluster densities. This is only a problem for $k$, and not for $\kappa$, because the former is a logarithmic measure, rendering it less sensitive to variations in $\mathcal{R}(m_1)$ than $\kappa(m_1)$.

We compute $k$ and $\kappa$ from our non-parametric reconstruction of the differential merger rate $\Gamma(m_1)$ inferred from the GW data. If the observations are consistent with a PISN mass gap that is populated by hierarchical mergers, both quantities might be expected to exhibit a maximum in the vicinity of $\tilde{m}$. For each posterior trace we evaluate $k$ and $\kappa$ over the range $m_1 > 30\,M_\odot$ and reconstruct the posterior distribution of the peak location, from which we extract the median and credible intervals. We obtain $m_*(k)=42.3^{+10.1}_{-4.3}\,M_\odot$ and $m_*(\kappa)=40.1^{+4.1}_{-2.2}\,M_\odot$. These values are consistent with the transitional mass $\tilde{m}$ inferred from the $\chi_{\rm eff}$ analysis, providing an independent test of the hypothesis that the observed transition is driven by the emergence of a distinct population above this mass.

\section*{Data Availability}
The data underlying this article can be downloaded
 at \href{https://zenodo.org/records/17148537}{https://zenodo.org/records/17148537}.

\section*{Code Availability}
The codes underlying this article can be found at 
\href{https://github.com/antoninifabio/spin-study-in-O4a}{\texttt{https://github.com/antoninifabio/spin-study-in-O4a}}.

\section*{Acknowledgements}
FA and FD are supported by the UK’s Science and Technology Facilities Council grants
ST/V005618/1 and UKRI2489. IMRS acknowledges support from the Science and Technology Facilities Council grant number ST/Y001990/1 and the Science and Technology Facilities Council Ernest Rutherford Fellowship grant number UKRI2423. DC acknowledges support from the Gordon and Betty Moore Foundation (Grant GBMF12341). This work was supported by a Leverhulme Trust International Professorship Grant (no.~LIP-2020-014). The work of Y.B.G.~was partly supported by a Simons Investigator Award to A.A.~Schekochihin and by the Science and Technology Facilities Council grant number ST/W00093/1.
{MG acknowledges financial support from the grants PID2024-155720NB-I00 and CEX2024-001451-M funded by MCIN/AEI/10.13039/501100011033 (State Agency for Research of the Spanish Ministry of Science and Innovation).}
This material is based upon work supported by
NSF’s LIGO Laboratory which is a major facility fully funded by
the National Science Foundation, as well as the Science and Technology Facilities Council (STFC) of the United Kingdom, the Max-Planck-Society (MPS), and the State of Niedersachsen/Germany for
support of the construction of Advanced LIGO and construction and
operation of the GEO600 detector. Additional support for Advanced
LIGO was provided by the Australian Research Council. Virgo is
funded, through the European Gravitational Observatory (EGO), by
the French Centre National de Recherche Scientifique (CNRS), the
Italian Istituto Nazionale di Fisica Nucleare (INFN) and the Dutch
Nikhef, with contributions by institutions from Belgium, Germany,
Greece, Hungary, Ireland, Japan, Monaco, Poland, Portugal, Spain.
KAGRA is supported by Ministry of Education, Culture, Sports, Science and Technology (MEXT), Japan Society for the Promotion of
Science (JSPS) in Japan; National Research Foundation (NRF) and
Ministry of Science and ICT (MSIT) in Korea; Academia Sinica
(AS) and National Science and Technology Council (NSTC) in Taiwan. The authors are grateful for computational resources provided
by Cardiff University and supported by STFC grant
ST/V005618/1. 

\section*{Author contribution statement}
F.A.,   I.M.R.S. and T.C. conceived the study. F.A. performed most of the  analysis, and wrote most of the manuscript. T.C. provided the original code on which the analysis code used in this manuscript is based on. F.D. calculated the $S_{300}$
 factor and assisted with incorporating the new GWTC-4 data into the analysis pipeline. Y.B.G. carried out the cluster simulations and the analysis presented in Method~\ref{PISNclusters}, and wrote the corresponding section. 
D.C., M.G., and M.P. contributed to the interpretation of the spin transition and its connection to the physics of PISN. 





\newpage

\begin{thebibliography}{82}
\ifx \bisbn   \undefined \def \bisbn  #1{ISBN #1}\fi
\ifx \binits  \undefined \def \binits#1{#1}\fi
\ifx \bauthor  \undefined \def \bauthor#1{#1}\fi
\ifx \batitle  \undefined \def \batitle#1{#1}\fi
\ifx \bjtitle  \undefined \def \bjtitle#1{#1}\fi
\ifx \bvolume  \undefined \def \bvolume#1{\textbf{#1}}\fi
\ifx \byear  \undefined \def \byear#1{#1}\fi
\ifx \bissue  \undefined \def \bissue#1{#1}\fi
\ifx \bfpage  \undefined \def \bfpage#1{#1}\fi
\ifx \blpage  \undefined \def \blpage #1{#1}\fi
\ifx \burl  \undefined \def \burl#1{\textsf{#1}}\fi
\ifx \doiurl  \undefined \def \doiurl#1{\url{https://doi.org/#1}}\fi
\ifx \betal  \undefined \def \betal{\textit{et al.}}\fi
\ifx \binstitute  \undefined \def \binstitute#1{#1}\fi
\ifx \binstitutionaled  \undefined \def \binstitutionaled#1{#1}\fi
\ifx \bctitle  \undefined \def \bctitle#1{#1}\fi
\ifx \beditor  \undefined \def \beditor#1{#1}\fi
\ifx \bpublisher  \undefined \def \bpublisher#1{#1}\fi
\ifx \bbtitle  \undefined \def \bbtitle#1{#1}\fi
\ifx \bedition  \undefined \def \bedition#1{#1}\fi
\ifx \bseriesno  \undefined \def \bseriesno#1{#1}\fi
\ifx \blocation  \undefined \def \blocation#1{#1}\fi
\ifx \bsertitle  \undefined \def \bsertitle#1{#1}\fi
\ifx \bsnm \undefined \def \bsnm#1{#1}\fi
\ifx \bsuffix \undefined \def \bsuffix#1{#1}\fi
\ifx \bparticle \undefined \def \bparticle#1{#1}\fi
\ifx \barticle \undefined \def \barticle#1{#1}\fi
\bibcommenthead
\ifx \bconfdate \undefined \def \bconfdate #1{#1}\fi
\ifx \botherref \undefined \def \botherref #1{#1}\fi
\ifx \bchapter \undefined \def \bchapter#1{#1}\fi
\ifx \bbook \undefined \def \bbook#1{#1}\fi
\ifx \bcomment \undefined \def \bcomment#1{#1}\fi
\ifx \oauthor \undefined \def \oauthor#1{#1}\fi
\ifx \citeauthoryear \undefined \def \citeauthoryear#1{#1}\fi
\ifx \endbibitem  \undefined \def \endbibitem {}\fi
\ifx \bconflocation  \undefined \def \bconflocation#1{#1}\fi
\ifx \arxivurl  \undefined \def \arxivurl#1{\textsf{#1}}\fi
\csname PreBibitemsHook\endcsname

\bibitem[\protect\citeauthoryear{{Abbott et al.}}{2019}]{2019PhRvX...9c1040A}
\begin{barticle}
\bauthor{\bsnm{{Abbott et al.}}}:
\batitle{GWTC-1: A Gravitational-Wave Transient Catalog of Compact Binary Mergers Observed by LIGO and Virgo during the First and Second Observing Runs}.
\bjtitle{Phys. Rev. X}
\bvolume{9}(\bissue{3}),
\bfpage{031040}
(\byear{2019})
\end{barticle}
\endbibitem

\bibitem[\protect\citeauthoryear{{LIGO Scientific Collaboration} and {Virgo Collaboration}}{2019}]{2019ApJ...882L..24A}
\begin{barticle}
\bauthor{\bsnm{{Abbott et al.}}}:
\batitle{Binary Black Hole Population Properties Inferred from the First and Second Observing Runs of Advanced LIGO and Advanced Virgo}.
\bjtitle{Astrophys. J. Lett.}
\bvolume{882}(\bissue{2}),
\bfpage{24}
(\byear{2019})
\end{barticle}
\endbibitem

\bibitem[\protect\citeauthoryear{{Abbott et al.}}{2021}]{Abbott:2020gyp}
\begin{barticle}
\bauthor{\bsnm{{Abbott et al.}}}:
\batitle{Population Properties of Compact Objects from the Second LIGO-Virgo Gravitational-Wave Transient Catalog}.
\bjtitle{Astrophys. J. Lett.}
\bvolume{913}(\bissue{1}),
\bfpage{7}
(\byear{2021})
\end{barticle}
\endbibitem

\bibitem[\protect\citeauthoryear{{Abbott et al.}}{2023}]{2021arXiv211103606T}
\begin{barticle}
\bauthor{\bsnm{{Abbott et al.}}}:
\batitle{GWTC-3: Compact Binary Coalescences Observed by LIGO and Virgo during the Second Part of the Third Observing Run}.
\bjtitle{Phys. Rev. X}
\bvolume{13}(\bissue{4}),
\bfpage{041039}
(\byear{2023})
\end{barticle}
\endbibitem

\bibitem[\protect\citeauthoryear{{Abbott et al.}}{2021}]{2021PhRvX..11b1053A}
\begin{barticle}
\bauthor{\bsnm{{Abbott et al.}}}:
\batitle{GWTC-2: Compact Binary Coalescences Observed by LIGO and Virgo during the First Half of the Third Observing Run}.
\bjtitle{Phys. Rev. X}
\bvolume{11}(\bissue{2}),
\bfpage{021053}
(\byear{2021})
\end{barticle}
\endbibitem

\bibitem[\protect\citeauthoryear{Spera et~al.}{2015}]{Spera2015a}
\begin{barticle}
\bauthor{\bsnm{Spera}, \binits{M.}},
\bauthor{\bsnm{Mapelli}, \binits{M.}},
\bauthor{\bsnm{Bressan}, \binits{A.}}:
\batitle{The mass spectrum of compact remnants from the PARSEC stellar evolution tracks}.
\bjtitle{Mon. Not. R. Astron. Soc.}
\bvolume{451}(\bissue{4}),
\bfpage{4086}--\blpage{4103}
(\byear{2015})
\end{barticle}
\endbibitem

\bibitem[\protect\citeauthoryear{Belczynski et~al.}{2016}]{belczynski2016}
\begin{barticle}
\bauthor{\bsnm{Belczynski}, \binits{K.}},
\bauthor{\bsnm{Holz}, \binits{D.E.}},
\bauthor{\bsnm{Bulik}, \binits{T.}},
\bauthor{\bsnm{O’Shaughnessy}, \binits{R.}}:
\batitle{The first gravitational-wave source from the isolated evolution of two stars in the 40–100 solar mass range}.
\bjtitle{Nature}
\bvolume{534}(\bissue{7608}),
\bfpage{512}--\blpage{515}
(\byear{2016})
\end{barticle}
\endbibitem

\bibitem[\protect\citeauthoryear{Stevenson et~al.}{2015}]{Stevenson2015}
\begin{barticle}
\bauthor{\bsnm{Stevenson}, \binits{S.}},
\bauthor{\bsnm{Ohme}, \binits{F.}},
\bauthor{\bsnm{Fairhurst}, \binits{S.}}:
\batitle{Distinguishing compact binary population synthesis models using gravitational wave observations of coalescing binary black holes}.
\bjtitle{Astrophys. J.}
\bvolume{810}(\bissue{1}),
\bfpage{58}
(\byear{2015})
\end{barticle}
\endbibitem

\bibitem[\protect\citeauthoryear{Marchant et~al.}{2016}]{Marchant2016}
\begin{barticle}
\bauthor{\bsnm{Marchant}, \binits{P.}},
\bauthor{\bsnm{Langer}, \binits{N.}},
\bauthor{\bsnm{Podsiadlowski}, \binits{P.}},
\bauthor{\bsnm{Tauris}, \binits{T.M.}},
\bauthor{\bsnm{Moriya}, \binits{T.J.}}:
\batitle{A new route towards merging massive black holes}.
\bjtitle{Astron. Astrophys.}
\bvolume{588},
\bfpage{A50}
(\byear{2016})
\end{barticle}
\endbibitem

\bibitem[\protect\citeauthoryear{{Woosley} and {Heger}}{2021}]{Woosley2021}
\begin{barticle}
\bauthor{\bsnm{Woosley}, \binits{S.E.}},
\bauthor{\bsnm{Heger}, \binits{A.}}:
\batitle{The Pair-instability Mass Gap for Black Holes}.
\bjtitle{Astrophys. J. Lett.}
\bvolume{912}(\bissue{2}),
\bfpage{31}
(\byear{2021})
\end{barticle}
\endbibitem
\bibitem[\protect\citeauthoryear{{Farrell et al.}}{2021}]{2021MNRAS.502L..40F}
\begin{barticle}
\bauthor{\bsnm{{Farrell et al.}}}:
\batitle{Is GW190521 the merger of black holes from the first stellar generations?}
\bjtitle{Mon. Not. R. Astron. Soc. Lett.}
\bvolume{502}(\bissue{1}),
\bfpage{40}--\blpage{44}
(\byear{2021})
\end{barticle}
\endbibitem

\bibitem[\protect\citeauthoryear{{Belczynski et al.}}{2020}]{2020ApJ...890..113B}
\begin{barticle}
\bauthor{\bsnm{{Belczynski et al.}}}:
\batitle{The Formation of a 70 M$_{{\ensuremath{\odot}}}$ Black Hole at High Metallicity}.
\bjtitle{Astrophys. J.}
\bvolume{890}(\bissue{2}),
\bfpage{113}
(\byear{2020})
\end{barticle}
\endbibitem

\bibitem[\protect\citeauthoryear{Woosley}{2017}]{Woosley2016}
\begin{barticle}
\bauthor{\bsnm{Woosley}, \binits{S.E.}}:
\batitle{Pulsational Pair-instability Supernovae}.
\bjtitle{Astrophys. J.}
\bvolume{836}(\bissue{2}),
\bfpage{244}
(\byear{2017})
\end{barticle}
\endbibitem

\bibitem[\protect\citeauthoryear{{Spera} and {Mapelli}}{2017}]{2017MNRAS.470.4739S}
\begin{barticle}
\bauthor{\bsnm{{Spera}}, \binits{M.}},
\bauthor{\bsnm{{Mapelli}}, \binits{M.}}:
\batitle{Very massive stars, pair-instability supernovae and intermediate-mass black holes with the SEVN code}.
\bjtitle{Mon. Not. R. Astron. Soc.}
\bvolume{470}(\bissue{4}),
\bfpage{4739}--\blpage{4749}
(\byear{2017})
\end{barticle}
\endbibitem

\bibitem[\protect\citeauthoryear{{Farmer} et~al.}{2020}]{2020ApJ...902L..36F}
\begin{barticle}
\bauthor{\bsnm{{Farmer}}, \binits{R.}},
\bauthor{\bsnm{{Renzo}}, \binits{M.}},
\bauthor{\bsnm{{de Mink}}, \binits{S.E.}},
\bauthor{\bsnm{{Fishbach}}, \binits{M.}},
\bauthor{\bsnm{{Justham}}, \binits{S.}}:
\batitle{Constraints from Gravitational-wave Detections of Binary Black Hole Mergers on the $^{12}$C({\ensuremath{\alpha}}, {\ensuremath{\gamma}})$^{16}$O Rate}.
\bjtitle{Astrophys. J. Lett.}
\bvolume{902}(\bissue{2}),
\bfpage{36}
(\byear{2020})
\end{barticle}
\endbibitem

\bibitem[\protect\citeauthoryear{{Hendriks} et~al.}{2023}]{2023MNRAS.526.4130H}
\begin{barticle}
\bauthor{\bsnm{{Hendriks}}, \binits{D.D.}},
\bauthor{\bsnm{{van Son}}, \binits{L.A.C.}},
\bauthor{\bsnm{{Renzo}}, \binits{M.}},
\bauthor{\bsnm{{Izzard}}, \binits{R.G.}},
\bauthor{\bsnm{{Farmer}}, \binits{R.}}:
\batitle{Pulsational pair-instability supernovae in gravitational-wave and electromagnetic transients}.
\bjtitle{Mon. Not. R. Astron. Soc.}
\bvolume{526}(\bissue{3}),
\bfpage{4130}--\blpage{4147}
(\byear{2023})
\end{barticle}
\endbibitem

\bibitem[\protect\citeauthoryear{{Farmer} et~al.}{2019}]{2019ApJ...887...53F}
\begin{barticle}
\bauthor{\bsnm{{Farmer}}, \binits{R.}},
\bauthor{\bsnm{{Renzo}}, \binits{M.}},
\bauthor{\bsnm{{de Mink}}, \binits{S.E.}},
\bauthor{\bsnm{{Marchant}}, \binits{P.}},
\bauthor{\bsnm{{Justham}}, \binits{S.}}:
\batitle{Mind the Gap: The Location of the Lower Edge of the Pair-instability Supernova Black Hole Mass Gap}.
\bjtitle{Astrophys. J.}
\bvolume{887}(\bissue{1}),
\bfpage{53}
(\byear{2019})
\end{barticle}
\endbibitem

\bibitem[\protect\citeauthoryear{{Leung} et~al.}{2019}]{2019ApJ...887...72L}
\begin{barticle}
\bauthor{\bsnm{{Leung}}, \binits{S.-C.}},
\bauthor{\bsnm{{Nomoto}}, \binits{K.}},
\bauthor{\bsnm{{Blinnikov}}, \binits{S.}}:
\batitle{Pulsational Pair-instability Supernovae. I. Pre-collapse Evolution and Pulsational Mass Ejection}.
\bjtitle{Astrophys. J.}
\bvolume{887}(\bissue{1}),
\bfpage{72}
(\byear{2019})
\end{barticle}
\endbibitem


\bibitem[\protect\citeauthoryear{{Edelman} et~al.}{2021}]{2021ApJ...913L..23E}
\begin{barticle}
\bauthor{\bsnm{{Edelman}}, \binits{B.}},
\bauthor{\bsnm{{Doctor}}, \binits{Z.}},
\bauthor{\bsnm{{Farr}}, \binits{B.}}:
\batitle{Poking Holes: Looking for Gaps in LIGO/Virgo's Black Hole Population}.
\bjtitle{Astrophys. J. Lett.}
\bvolume{913}(\bissue{2}),
\bfpage{23}
(\byear{2021})
\end{barticle}
\endbibitem

\bibitem[\protect\citeauthoryear{{Abbott et al.}}{2023}]{LVKCollab2023}
\begin{barticle}
\bauthor{\bsnm{{Abbott et al.}}}:
\batitle{Population of Merging Compact Binaries Inferred Using Gravitational Waves through GWTC-3}.
\bjtitle{Phys. Rev. X}
\bvolume{13}(\bissue{1}),
\bfpage{011048}
(\byear{2023})
\end{barticle}
\endbibitem

\bibitem[\protect\citeauthoryear{{Ray} et~al.}{2023}]{ray_nonparametric_2023}
\begin{barticle}
\bauthor{\bsnm{{Ray}}, \binits{A.}},
\bauthor{\bsnm{{Hernandez}}, \binits{I.M.}},
\bauthor{\bsnm{{Mohite}}, \binits{S.}},
\bauthor{\bsnm{{Creighton}}, \binits{J.}},
\bauthor{\bsnm{{Kapadia}}, \binits{S.}}:
\batitle{Nonparametric Inference of the Population of Compact Binaries from Gravitational-wave Observations Using Binned Gaussian Processes}.
\bjtitle{Astrophys. J.}
\bvolume{957}(\bissue{1}),
\bfpage{37}
(\byear{2023})
\end{barticle}
\endbibitem

\bibitem[\protect\citeauthoryear{{Callister} and {Farr}}{2024}]{2024PhRvX..14b1005C}
\begin{barticle}
\bauthor{\bsnm{{Callister}}, \binits{T.A.}},
\bauthor{\bsnm{{Farr}}, \binits{W.M.}}:
\batitle{Parameter-Free Tour of the Binary Black Hole Population}.
\bjtitle{Phys. Rev. X}
\bvolume{14}(\bissue{2}),
\bfpage{021005}
(\byear{2024})
\end{barticle}
\endbibitem

\bibitem[\protect\citeauthoryear{{Afroz} and {Mukherjee}}{2025}]{2025PhRvD.112b3531A}
\begin{barticle}
\bauthor{\bsnm{{Afroz}}, \binits{S.}},
\bauthor{\bsnm{{Mukherjee}}, \binits{S.}}:
\batitle{Phase space of binary black holes from gravitational wave observations to unveil its formation history}.
\bjtitle{Phys. Rev. D}
\bvolume{112}(\bissue{2}),
\bfpage{023531}
(\byear{2025})
\end{barticle}
\endbibitem

\bibitem[\protect\citeauthoryear{{Olejak} et~al.}{2022}]{2022MNRAS.516.2252O}
\begin{barticle}
\bauthor{\bsnm{{Olejak}}, \binits{A.}},
\bauthor{\bsnm{{Fryer}}, \binits{C.L.}},
\bauthor{\bsnm{{Belczynski}}, \binits{K.}},
\bauthor{\bsnm{{Baibhav}}, \binits{V.}}:
\batitle{The role of supernova convection for the lower mass gap in the isolated binary formation of gravitational wave sources}.
\bjtitle{Mon. Not. R. Astron. Soc.}
\bvolume{516}(\bissue{2}),
\bfpage{2252}--\blpage{2271}
(\byear{2022})
\end{barticle}
\endbibitem

\bibitem[\protect\citeauthoryear{{Winch} et~al.}{2024}]{2024MNRAS.529.2980W}
\begin{barticle}
\bauthor{\bsnm{{Winch}}, \binits{E.R.J.}},
\bauthor{\bsnm{{Vink}}, \binits{J.S.}},
\bauthor{\bsnm{{Higgins}}, \binits{E.R.}},
\bauthor{\bsnm{{Sabhahitf}}, \binits{G.N.}}:
\batitle{Predicting the heaviest black holes below the pair instability gap}.
\bjtitle{Mon. Not. R. Astron. Soc.}
\bvolume{529}(\bissue{3}),
\bfpage{2980}--\blpage{3002}
(\byear{2024})
\end{barticle}
\endbibitem

\bibitem[\protect\citeauthoryear{{Costa et al.}}{2021}]{2021MNRAS.501.4514C}
\begin{barticle}
\bauthor{\bsnm{{Costa et al.}}}:
\batitle{Formation of GW190521 from stellar evolution: the impact of the hydrogen-rich envelope, dredge-up, and $^{12}$C({\ensuremath{\alpha}}, {\ensuremath{\gamma}})$^{16}$O rate on the pair-instability black hole mass gap}.
\bjtitle{Mon. Not. R. Astron. Soc.}
\bvolume{501}(\bissue{3}),
\bfpage{4514}--\blpage{4533}
(\byear{2021})
\end{barticle}
\endbibitem

\bibitem[\protect\citeauthoryear{Rezzolla et~al.}{2008}]{PhysRevD.78.044002}
\begin{barticle}
\bauthor{\bsnm{Rezzolla et al.}}:
\batitle{Final spin from the coalescence of two black holes}.
\bjtitle{Phys. Rev. D}
\bvolume{78},
\bfpage{044002}
(\byear{2008})
\end{barticle}
\endbibitem

\bibitem[\protect\citeauthoryear{Campanelli et~al.}{2006}]{PhysRevLett.96.111101}
\begin{barticle}
\bauthor{\bsnm{Campanelli}, \binits{M.}},
\bauthor{\bsnm{Lousto}, \binits{C.O.}},
\bauthor{\bsnm{Marronetti}, \binits{P.}},
\bauthor{\bsnm{Zlochower}, \binits{Y.}}:
\batitle{Accurate evolutions of orbiting black-hole binaries without excision}.
\bjtitle{Phys. Rev. Lett.}
\bvolume{96},
\bfpage{111101}
(\byear{2006})
\end{barticle}
\endbibitem

\bibitem[\protect\citeauthoryear{Baker et~al.}{2006}]{PhysRevLett.96.111102}
\begin{barticle}
\bauthor{\bsnm{Baker}, \binits{J.G.}},
\bauthor{\bsnm{Centrella}, \binits{J.}},
\bauthor{\bsnm{Choi}, \binits{D.-I.}},
\bauthor{\bsnm{Koppitz}, \binits{M.}},
\bauthor{\bsnm{Meter}, \binits{J.}}:
\batitle{Gravitational-wave extraction from an inspiraling configuration of merging black holes}.
\bjtitle{Phys. Rev. Lett.}
\bvolume{96},
\bfpage{111102}
(\byear{2006})
\end{barticle}
\endbibitem

\bibitem[\protect\citeauthoryear{{Chattopadhyay} et~al.}{2023}]{2023MNRAS.526.4908C}
\begin{barticle}
\bauthor{\bsnm{{Chattopadhyay}}, \binits{D.}},
\bauthor{\bsnm{{Stegmann}}, \binits{J.}},
\bauthor{\bsnm{{Antonini}}, \binits{F.}},
\bauthor{\bsnm{{Barber}}, \binits{J.}},
\bauthor{\bsnm{{Romero-Shaw}}, \binits{I.M.}}:
\batitle{Double black hole mergers in nuclear star clusters: eccentricities, spins, masses, and the growth of massive seeds}.
\bjtitle{Mon. Not. R. Astron. Soc.}
\bvolume{526}(\bissue{4}),
\bfpage{4908}--\blpage{4928}
(\byear{2023})
\end{barticle}
\endbibitem

\bibitem[\protect\citeauthoryear{{O'Leary} et~al.}{2006}]{2006ApJ...637..937O}
\begin{barticle}
\bauthor{\bsnm{{O'Leary}}, \binits{R.M.}},
\bauthor{\bsnm{{Rasio}}, \binits{F.A.}},
\bauthor{\bsnm{{Fregeau}}, \binits{J.M.}},
\bauthor{\bsnm{{Ivanova}}, \binits{N.}},
\bauthor{\bsnm{{O'Shaughnessy}}, \binits{R.}}:
\batitle{Binary Mergers and Growth of Black Holes in Dense Star Clusters}.
\bjtitle{Astrophys. J.}
\bvolume{637}(\bissue{2}),
\bfpage{937}--\blpage{951}
(\byear{2006})
\end{barticle}
\endbibitem

\bibitem[\protect\citeauthoryear{{Antonini} and {Rasio}}{2016}]{2016ApJ...831..187A}
\begin{barticle}
\bauthor{\bsnm{{Antonini}}, \binits{F.}},
\bauthor{\bsnm{{Rasio}}, \binits{F.A.}}:
\batitle{Merging Black Hole Binaries in Galactic Nuclei: Implications for Advanced-LIGO Detections}.
\bjtitle{Astrophys. J.}
\bvolume{831}(\bissue{2}),
\bfpage{187}
(\byear{2016})
\end{barticle}
\endbibitem

\bibitem[\protect\citeauthoryear{Rodriguez et~al.}{2015}]{Rodriguez2015a}
\begin{barticle}
\bauthor{\bsnm{Rodriguez et al.}}:
\batitle{Binary Black Hole Mergers from Globular Clusters: Implications for Advanced LIGO}.
\bjtitle{Phys. Rev. Lett.}
\bvolume{115}(\bissue{5}),
\bfpage{051101}
(\byear{2015})
\end{barticle}
\endbibitem

\bibitem[\protect\citeauthoryear{{Racine}}{2008}]{2008PhRvD..78d4021R}
\begin{barticle}
\bauthor{\bsnm{{Racine}}, \binits{{\'E}.}}:
\batitle{Analysis of spin precession in binary black hole systems including quadrupole-monopole interaction}.
\bjtitle{Phys. Rev. D}
\bvolume{78}(\bissue{4}),
\bfpage{044021}
(\byear{2008})
\end{barticle}
\endbibitem

\bibitem[\protect\citeauthoryear{Ajith et~al.}{2011}]{PhysRevLett.106.241101}
\begin{barticle}
\bauthor{\bsnm{Ajith et al.}}:
\batitle{Inspiral-merger-ringdown waveforms for black-hole binaries with nonprecessing spins}.
\bjtitle{Phys. Rev. Lett.}
\bvolume{106},
\bfpage{241101}
(\byear{2011})
\end{barticle}
\endbibitem

\bibitem[\protect\citeauthoryear{{Antonini} et~al.}{2025}]{2025PhRvL.134a1401A}
\begin{barticle}
\bauthor{\bsnm{{Antonini}}, \binits{F.}},
\bauthor{\bsnm{{Romero-Shaw}}, \binits{I.M.}},
\bauthor{\bsnm{{Callister}}, \binits{T.}}:
\batitle{Star Cluster Population of High Mass Black Hole Mergers in Gravitational Wave Data}.
\bjtitle{Phys. Rev. Lett.}
\bvolume{134}(\bissue{1}),
\bfpage{011401}
(\byear{2025})
\end{barticle}
\endbibitem

\bibitem[\protect\citeauthoryear{{Rodriguez et al.}}{2019}]{2019PhRvD.100d3027R}
\begin{barticle}
\bauthor{\bsnm{{Rodriguez et al.}}}:
\batitle{Black holes: The next generation---repeated mergers in dense star clusters and their gravitational-wave properties}.
\bjtitle{Phys. Rev. D}
\bvolume{100}(\bissue{4}),
\bfpage{043027}
(\byear{2019})
\end{barticle}
\endbibitem

\bibitem[\protect\citeauthoryear{{The LIGO Scientific Collaboration} et~al.}{2025a}]{LVKpop_inprep}
\begin{botherref}
\oauthor{\bsnm{Abac, A.~G. et al. 2025}}:
{GWTC-4.0: Population Properties of Merging Compact Binaries}.
arXiv e-prints,
2508--18083
(2025)
\end{botherref}
\endbibitem

\bibitem[\protect\citeauthoryear{{The LIGO Scientific Collaboration} et~al.}{2025b}]{LVKcat_inprep}
\begin{botherref}
\oauthor{\bsnm{Abac, A.~G. et al. 2025}}:
{GWTC-4.0: Updating the Gravitational-Wave Transient Catalog with Observations from the First Part of the Fourth LIGO-Virgo-KAGRA Observing Run}.
arXiv e-prints,
2508--18082
(2025)
\end{botherref}
\endbibitem

\bibitem[\protect\citeauthoryear{Collaboration et~al.}{2025}]{LVK_GWTC4_2025}
\begin{botherref}
\oauthor{\bsnm{LIGO-Virgo-KAGRA Collaboration}}:
GWTC-4: Compact Binary Coalescences Observed by LIGO and Virgo During the O3b Observing Run.
Zenodo
(2025).
\url{https://zenodo.org/records/16053484}
\end{botherref}
\endbibitem

\bibitem[\protect\citeauthoryear{{Antonini} et~al.}{2023}]{2023MNRAS.522..466A}
\begin{barticle}
\bauthor{\bsnm{{Antonini}}, \binits{F.}},
\bauthor{\bsnm{{Gieles}}, \binits{M.}},
\bauthor{\bsnm{{Dosopoulou}}, \binits{F.}},
\bauthor{\bsnm{{Chattopadhyay}}, \binits{D.}}:
\batitle{Coalescing black hole binaries from globular clusters: mass distributions and comparison to gravitational wave data from GWTC-3}.
\bjtitle{Mon. Not. R. Astron. Soc.}
\bvolume{522}(\bissue{1}),
\bfpage{466}--\blpage{476}
(\byear{2023})
\end{barticle}
\endbibitem

\bibitem[\protect\citeauthoryear{{Tiwari}}{2024}]{2024MNRAS.527..298T}
\begin{barticle}
\bauthor{\bsnm{{Tiwari}}, \binits{V.}}:
\batitle{What's in a binary black hole's mass parameter?}
\bjtitle{Mon. Not. R. Astron. Soc.}
\bvolume{527}(\bissue{1}),
\bfpage{298}--\blpage{306}
(\byear{2024})
\end{barticle}
\endbibitem

\bibitem[\protect\citeauthoryear{{Abac et al.}}{2025}]{2025ApJ...993L..21A}
\begin{barticle}
\bauthor{\bsnm{{Abac et al.}}}:
\batitle{GW241011 and GW241110: Exploring Binary Formation and Fundamental Physics with Asymmetric, High-spin Black Hole Coalescences}.
\bjtitle{Astrophys. J. Lett.}
\bvolume{993}(\bissue{1}),
\bfpage{21}
(\byear{2025})
\end{barticle}
\endbibitem
\bibitem[\protect\citeauthoryear{{Tiwari} and {Fairhurst}}{2021}]{2021ApJ...913L..19T}
\begin{barticle}
\bauthor{\bsnm{{Tiwari}}, \binits{V.}},
\bauthor{\bsnm{{Fairhurst}}, \binits{S.}}:
\batitle{The Emergence of Structure in the Binary Black Hole Mass Distribution}.
\bjtitle{Astrophys. J. Lett.}
\bvolume{913}(\bissue{2}),
\bfpage{19}
(\byear{2021})
\end{barticle}
\endbibitem

\bibitem[\protect\citeauthoryear{{Wang et al.}}{2022}]{2022ApJ...941L..39W}
\begin{barticle}
\bauthor{\bsnm{{Wang et al.}}}:
\batitle{Potential Subpopulations and Assembling Tendency of the Merging Black Holes}.
\bjtitle{Astrophys. J. Lett.}
\bvolume{941}(\bissue{2}),
\bfpage{39}
(\byear{2022})
\end{barticle}
\endbibitem

\bibitem[\protect\citeauthoryear{{Li} et~al.}{2024}]{2023arXiv230302973L}
\begin{barticle}
\bauthor{\bsnm{{Li}}, \binits{Y.-J.}},
\bauthor{\bsnm{{Wang}}, \binits{Y.-Z.}},
\bauthor{\bsnm{{Tang}}, \binits{S.-P.}},
\bauthor{\bsnm{{Fan}}, \binits{Y.-Z.}}:
\batitle{Resolving the Stellar-Collapse and Hierarchical-Merger Origins of the Coalescing Black Holes}.
\bjtitle{Phys. Rev. Lett.}
\bvolume{133}(\bissue{5}),
\bfpage{051401}
(\byear{2024})
\end{barticle}
\endbibitem

\bibitem[\protect\citeauthoryear{{Antonini} et~al.}{2025}]{2025arXiv250609154A}
\begin{botherref}
\oauthor{\bsnm{{Antonini}}, \binits{F.}},
\oauthor{\bsnm{{Callister}}, \binits{T.}},
\oauthor{\bsnm{{Dosopoulou}}, \binits{F.}},
\oauthor{\bsnm{{Romero-Shaw}}, \binits{I.}},
\oauthor{\bsnm{{Chattopadhyay}}, \binits{D.}}:
{Inferring the pair-instability mass gap from gravitational wave data using flexible models}.
arXiv e-prints,
2506--09154
(2025)
\end{botherref}
\endbibitem

\bibitem[\protect\citeauthoryear{{Sadiq} et~al.}{2025}]{2025arXiv250602250S}
\begin{botherref}
\oauthor{\bsnm{{Sadiq}}, \binits{J.}},
\oauthor{\bsnm{{Dent}}, \binits{T.}},
\oauthor{\bsnm{{Lorenzo-Medina}}, \binits{A.}}:
{Seeking Spinning Subpopulations of Black Hole Binaries via Iterative Density Estimation}.
arXiv e-prints,
2506--02250
(2025)
\end{botherref}
\endbibitem

\bibitem[\protect\citeauthoryear{{Maga{\~n}a Hernandez} and {Palmese}}{2025}]{2025arXiv250819208M}
\begin{botherref}
\oauthor{\bsnm{{Maga{\~n}a Hernandez}}, \binits{I.}},
\oauthor{\bsnm{{Palmese}}, \binits{A.}}:
{Astrophysics informed Gaussian processes for gravitational-wave populations: Evidence for the onset of the pair-instability supernova mass gap}.
arXiv e-prints,
2508--19208
(2025)
\end{botherref}
\endbibitem

\bibitem[\protect\citeauthoryear{{Tong et al.}}{2025}]{2025arXiv250904151T}
\begin{botherref}
\oauthor{\bsnm{{Tong et al.}}}:
{Evidence of the pair instability gap in the distribution of black hole masses}.
arXiv e-prints,
2509--04151
(2025)
\end{botherref}
\endbibitem

\bibitem[\protect\citeauthoryear{{Liu} and {Lai}}{2021}]{2021MNRAS.502.2049L}
\begin{barticle}
\bauthor{\bsnm{{Liu}}, \binits{B.}},
\bauthor{\bsnm{{Lai}}, \binits{D.}}:
\batitle{Hierarchical black hole mergers in multiple systems: constrain the formation of GW190412-, GW190814-, and GW190521-like events}.
\bjtitle{Mon. Not. R. Astron. Soc.}
\bvolume{502}(\bissue{2}),
\bfpage{2049}--\blpage{2064}
(\byear{2021})
\end{barticle}
\endbibitem

\bibitem[\protect\citeauthoryear{{Sallaska et al.}}{2013}]{2013ApJS..207...18S}
\begin{barticle}
\bauthor{\bsnm{{Sallaska et al.}}}:
\batitle{STARLIB: A Next-generation Reaction-rate Library for Nuclear Astrophysics}.
\bjtitle{Astrophys. J. Suppl. Ser.}
\bvolume{207}(\bissue{1}),
\bfpage{18}
(\byear{2013})
\end{barticle}
\endbibitem


\bibitem[\protect\citeauthoryear{{Golomb} et~al.}{2024}]{2024ApJ...976..121G}
\begin{barticle}
\bauthor{\bsnm{{Golomb}}, \binits{J.}},
\bauthor{\bsnm{{Isi}}, \binits{M.}},
\bauthor{\bsnm{{Farr}}, \binits{W.M.}}:
\batitle{Physical Models for the Astrophysical Population of Black Holes: Application to the Bump in the Mass Distribution of Gravitational-wave Sources}.
\bjtitle{Astrophys. J.}
\bvolume{976}(\bissue{1}),
\bfpage{121}
(\byear{2024})
\end{barticle}
\endbibitem

\bibitem[\protect\citeauthoryear{{An} et~al.}{2015}]{An2015}
\begin{barticle}
\bauthor{\bsnm{{An}}, \binits{Z.}},
\bauthor{\bsnm{{Ma}}, \binits{Z.-Y.}},
\bauthor{\bsnm{{Yuan}}, \binits{C.-L.}},
\bauthor{\bsnm{{Meng}}, \binits{J.}}:
\batitle{New analysis of the $^{12}$C($\alpha,\gamma$)$^{16}$O reaction cross section at astrophysical energies}.
\bjtitle{Phys. Rev. C}
\bvolume{92}(\bissue{1}),
\bfpage{015802}
(\byear{2015})
\end{barticle}
\endbibitem

\bibitem[\protect\citeauthoryear{{de Boer et al.}}{2025}]{deBoer2025}
\begin{barticle}
\bauthor{\bsnm{{de Boer et al.}}}:
\batitle{The $^{12}$C($\alpha,\gamma$)$^{16}$O reaction revisited: a new R-matrix evaluation}.
\bjtitle{Eur. Phys. J. A}
\bvolume{61}(\bissue{2}),
\bfpage{37}
(\byear{2025})
\end{barticle}
\endbibitem

\bibitem[\protect\citeauthoryear{{Shen et al.}}{2023}]{Shen2023}
\begin{barticle}
\bauthor{\bsnm{{Shen et al.}}}:
\batitle{New Determination of the $^{12}$C({\ensuremath{\alpha}}, {\ensuremath{\gamma}})$^{16}$O Reaction Rate and Its Impact on the Black-hole Mass Gap}.
\bjtitle{Astrophys. J.}
\bvolume{945}(\bissue{1}),
\bfpage{41}
(\byear{2023})
\end{barticle}
\endbibitem

\bibitem[\protect\citeauthoryear{{Salaris et al.}}{1997}]{1997ApJ...486..413S}
\begin{barticle}
\bauthor{\bsnm{{Salaris et al.}}}:
\batitle{The Cooling of CO White Dwarfs: Influence of the Internal Chemical Distribution}.
\bjtitle{Astrophys. J.}
\bvolume{486}(\bissue{1}),
\bfpage{413}--\blpage{419}
(\byear{1997})
\end{barticle}
\endbibitem

\bibitem[\protect\citeauthoryear{{Woosley}}{2019}]{2019ApJ...878...49W}
\begin{barticle}
\bauthor{\bsnm{{Woosley}}, \binits{S.E.}}:
\batitle{The Evolution of Massive Helium Stars, Including Mass Loss}.
\bjtitle{Astrophys. J.}
\bvolume{878}(\bissue{1}),
\bfpage{49}
(\byear{2019})
\end{barticle}
\endbibitem

\bibitem[\protect\citeauthoryear{{Boothroyd} and {Sackmann}}{1988}]{1988ApJ...328..653B}
\begin{barticle}
\bauthor{\bsnm{{Boothroyd}}, \binits{A.I.}},
\bauthor{\bsnm{{Sackmann}}, \binits{I.-J.}}:
\batitle{Low-Mass Stars. III. Low-Mass Stars with Steady Mass Loss: Up to the Asymptotic Giant Branch and through the Final Thermal Pulses}.
\bjtitle{Astrophys. J.}
\bvolume{328},
\bfpage{653}
(\byear{1988})
\end{barticle}
\endbibitem

\bibitem[\protect\citeauthoryear{{{\"O}berg} et~al.}{2011}]{2011ApJ...743L..16O}
\begin{barticle}
\bauthor{\bsnm{{{\"O}berg}}, \binits{K.I.}},
\bauthor{\bsnm{{Murray-Clay}}, \binits{R.}},
\bauthor{\bsnm{{Bergin}}, \binits{E.A.}}:
\batitle{The Effects of Snowlines on C/O in Planetary Atmospheres}.
\bjtitle{Astrophys. J. Lett.}
\bvolume{743}(\bissue{1}),
\bfpage{16}
(\byear{2011})
\end{barticle}
\endbibitem

\bibitem[\protect\citeauthoryear{{Tiwari}}{2022}]{2022ApJ...928..155T}
\begin{barticle}
\bauthor{\bsnm{{Tiwari}}, \binits{V.}}:
\batitle{Exploring Features in the Binary Black Hole Population}.
\bjtitle{Astrophys. J.}
\bvolume{928}(\bissue{2}),
\bfpage{155}
(\byear{2022})
\end{barticle}
\endbibitem

\bibitem[\protect\citeauthoryear{Vink et~al.}{2001}]{Vink2001}
\begin{barticle}
\bauthor{\bsnm{Vink}, \binits{J.S.}},
\bauthor{\bsnm{Koter}, \binits{A.}},
\bauthor{\bsnm{Lamers}, \binits{H.J.G.L.M.}}:
\batitle{Mass-loss predictions for O and B stars as a function of metallicity}.
\bjtitle{Astron. Astrophys.}
\bvolume{369}(\bissue{2}),
\bfpage{574}--\blpage{588}
(\byear{2001})
\end{barticle}
\endbibitem

\bibitem[\protect\citeauthoryear{{Bartos} et~al.}{2017}]{Bartos2016}
\begin{barticle}
\bauthor{\bsnm{{Bartos}}, \binits{I.}},
\bauthor{\bsnm{{Kocsis}}, \binits{B.}},
\bauthor{\bsnm{{Haiman}}, \binits{Z.}},
\bauthor{\bsnm{{M{\'a}rka}}, \binits{S.}}:
\batitle{Rapid and Bright Stellar-mass Binary Black Hole Mergers in Active Galactic Nuclei}.
\bjtitle{Astrophys. J.}
\bvolume{835}(\bissue{2}),
\bfpage{165}
(\byear{2017})
\end{barticle}
\endbibitem

\bibitem[\protect\citeauthoryear{Ray and Kalogera}{2025}]{2025arXiv251018867R}
\begin{botherref}
\oauthor{\bsnm{Ray}, A.} and \oauthor{\bsnm{Kalogera}, V.}:
Reexamining Evidence of a Pair-Instability Mass Gap in the Binary Black Hole Population.
arXiv e-prints
(2025).
\href{https://doi.org/10.48550/arXiv.2510.18867}{doi:10.48550/arXiv.2510.18867}
\end{botherref}
\endbibitem


@ARTICLE{2025arXiv251018867R,
  author       = {{Ray}, Anarya and {Kalogera}, Vicky},
  title        = {Reexamining Evidence of a Pair-Instability Mass Gap in the Binary Black Hole Population},
  journal      = {arXiv e-prints},
  year         = 2025,
  month        = oct,
  eid          = {arXiv:2510.18867},
  pages        = {arXiv:2510.18867},
  doi          = {10.48550/arXiv.2510.18867},
  archivePrefix= {arXiv},
  eprint       = {2510.18867},
  primaryClass = {astro-ph.HE},
  adsurl       = {https://ui.adsabs.harvard.edu/abs/2025arXiv251018867R},
  adsnote      = {Provided by the SAO/NASA Astrophysics Data System}
}

\bibitem[\protect\citeauthoryear{Collaboration et~al.}{2023}]{LVK_GWTC3_2023}
\begin{botherref}
\oauthor{\bsnm{LIGO-Virgo-KAGRA Collaboration}}:
GWTC-3: Compact Binary Coalescences Observed by LIGO and Virgo During the Second Part of the Third Observing Run --- Data behind the figures.
Zenodo
(2023).
\url{https://zenodo.org/records/7997424}
\end{botherref}
\endbibitem

\bibitem[\protect\citeauthoryear{{Abbott et al.}}{2024}]{2024PhRvD.109b2001A}
\begin{barticle}
\bauthor{\bsnm{{Abbott et al.}}}:
\batitle{GWTC-2.1: Deep extended catalog of compact binary coalescences observed by LIGO and Virgo during the first half of the third observing run}.
\bjtitle{Phys. Rev. D}
\bvolume{109}(\bissue{2}),
\bfpage{022001}
(\byear{2024})
\end{barticle}
\endbibitem

\bibitem[\protect\citeauthoryear{{Varma et al.}}{2019}]{2019PhRvR...1c3015V}
\begin{barticle}
\bauthor{\bsnm{{Varma et al.}}}:
\batitle{Surrogate models for precessing binary black hole simulations with unequal masses}.
\bjtitle{Phys. Rev. Res.}
\bvolume{1}(\bissue{3}),
\bfpage{033015}
(\byear{2019})
\end{barticle}
\endbibitem

\bibitem[\protect\citeauthoryear{{Buikema et al.}}{2020}]{2020PhRvD.102f2003B}
\begin{barticle}
\bauthor{\bsnm{{Buikema et al.}}}:
\batitle{Sensitivity and performance of the Advanced LIGO detectors in the third observing run}.
\bjtitle{Phys. Rev. D}
\bvolume{102}(\bissue{6}),
\bfpage{062003}
(\byear{2020})
\end{barticle}
\endbibitem

\bibitem[\protect\citeauthoryear{{Capote et al.}}{2025}]{2025PhRvD.111f2002C}
\begin{barticle}
\bauthor{\bsnm{{Capote et al.}}}:
\batitle{Advanced LIGO detector performance in the fourth observing run}.
\bjtitle{Phys. Rev. D}
\bvolume{111}(\bissue{6}),
\bfpage{062002}
(\byear{2025})
\end{barticle}
\endbibitem

\bibitem[\protect\citeauthoryear{{Soni et al.}}{2025}]{2025CQGra..42h5016S}
\begin{barticle}
\bauthor{\bsnm{{Soni et al.}}}:
\batitle{LIGO Detector Characterization in the first half of the fourth Observing run}.
\bjtitle{Class. Quantum Gravity}
\bvolume{42}(\bissue{8}),
\bfpage{085016}
(\byear{2025})
\end{barticle}
\endbibitem

\bibitem[\protect\citeauthoryear{{Novikov et al.}}{2025}]{2025Natur.643..955N}
\begin{barticle}
\bauthor{\bsnm{{Novikov et al.}}}:
\batitle{Hybrid quantum network for sensing in the acoustic frequency range}.
\bjtitle{Nature}
\bvolume{643}(\bissue{8073}),
\bfpage{955}--\blpage{960}
(\byear{2025})
\end{barticle}
\endbibitem

\bibitem[\protect\citeauthoryear{Ganapathy et~al.}{2023}]{PhysRevX.13.041021}
\begin{barticle}
\bauthor{\bsnm{Ganapathy et al.}}:
\batitle{Broadband quantum enhancement of the ligo detectors with frequency-dependent squeezing}.
\bjtitle{Phys. Rev. X}
\bvolume{13},
\bfpage{041021}
(\byear{2023})
\end{barticle}
\endbibitem

\bibitem[\protect\citeauthoryear{{Jia et al.}}{2024}]{2024Sci...385.1318J}
\begin{barticle}
\bauthor{\bsnm{{Jia et al.}}}:
\batitle{Squeezing the quantum noise of a gravitational-wave detector below the standard quantum limit}.
\bjtitle{Science}
\bvolume{385}(\bissue{6715}),
\bfpage{1318}--\blpage{1321}
(\byear{2024})
\end{barticle}
\endbibitem

\bibitem[\protect\citeauthoryear{{The LIGO Scientific Collaboration, Virgo Collaboration, and KAGRA Collaboration}}{2021}]{injections}
\begin{botherref}
\oauthor{\bsnm{{The LIGO Scientific Collaboration, Virgo Collaboration, and KAGRA Collaboration}}}:
{GWTC-3: Compact Binary Coalescences Observed by LIGO and Virgo During the Second Part of the Third Observing Run --- O1+O2+O3 Search Sensitivity Estimates}.
\url{https://doi.org/10.5281/zenodo.5636816}
\end{botherref}
\endbibitem

\bibitem[\protect\citeauthoryear{{Callister} et~al.}{2022}]{2022ApJ...937L..13C}
\begin{barticle}
\bauthor{\bsnm{{Callister}}, \binits{T.A.}},
\bauthor{\bsnm{{Miller}}, \binits{S.J.}},
\bauthor{\bsnm{{Chatziioannou}}, \binits{K.}},
\bauthor{\bsnm{{Farr}}, \binits{W.M.}}:
\batitle{No Evidence that the Majority of Black Holes in Binaries Have Zero Spin}.
\bjtitle{Astrophys. J. Lett.}
\bvolume{937}(\bissue{1}),
\bfpage{13}
(\byear{2022})
\end{barticle}
\endbibitem

\bibitem[\protect\citeauthoryear{Fishbach et~al.}{2018}]{Fishbach_2018}
\begin{barticle}
\bauthor{\bsnm{Fishbach}, \binits{M.}},
\bauthor{\bsnm{Holz}, \binits{D.E.}},
\bauthor{\bsnm{Farr}, \binits{W.M.}}:
\batitle{Does the black hole merger rate evolve with redshift?}
\bjtitle{Astrophys. J. Lett.}
\bvolume{863}(\bissue{2}),
\bfpage{41}
(\byear{2018})
\end{barticle}
\endbibitem

\bibitem[\protect\citeauthoryear{Callister et~al.}{2020}]{Callister_2020}
\begin{barticle}
\bauthor{\bsnm{Callister}, \binits{T.}},
\bauthor{\bsnm{Fishbach}, \binits{M.}},
\bauthor{\bsnm{Holz}, \binits{D.E.}},
\bauthor{\bsnm{Farr}, \binits{W.M.}}:
\batitle{Shouts and murmurs: Combining individual gravitational-wave sources with the stochastic background to measure the history of binary black hole mergers}.
\bjtitle{Astrophys. J. Lett.}
\bvolume{896}(\bissue{2}),
\bfpage{32}
(\byear{2020})
\end{barticle}
\endbibitem

\bibitem[\protect\citeauthoryear{Gelman et~al.}{2013}]{gelman2013bda3}
\begin{bbook}
\bauthor{\bsnm{Gelman et al.}}:
\bbtitle{Bayesian Data Analysis},
\bedition{3rd} edn.
\bpublisher{CRC Press}, \blocation{???}
(\byear{2013})
\end{bbook}
\endbibitem

\bibitem[\protect\citeauthoryear{{Galaudage} and {Lamberts}}{2025}]{2025A&A...694A.186G}
\begin{barticle}
\bauthor{\bsnm{{Galaudage}}, \binits{S.}},
\bauthor{\bsnm{{Lamberts}}, \binits{A.}}:
\batitle{Compactness peaks: An astrophysical interpretation of the mass distribution of merging binary black holes}.
\bjtitle{Astron. Astrophys.}
\bvolume{694},
\bfpage{186}
(\byear{2025})
\end{barticle}
\endbibitem



\bibitem[\protect\citeauthoryear{{Fraley}}{1968}]{1968Ap&SS...2...96F}
\begin{barticle}
\bauthor{\bsnm{{Fraley}}, \binits{G.S.}}:
\batitle{Supernovae Explosions Induced by Pair-Production Instability}.
\bjtitle{Astrophys. Space Sci.}
\bvolume{2}(\bissue{1}),
\bfpage{96}--\blpage{114}
(\byear{1968})
\end{barticle}
\endbibitem


\bibitem[\protect\citeauthoryear{{Vink} et~al.}{2021}]{2021MNRAS.504..146V}
\begin{barticle}
\bauthor{\bsnm{{Vink}}, \binits{J.S.}},
\bauthor{\bsnm{{Higgins}}, \binits{E.R.}},
\bauthor{\bsnm{{Sander}}, \binits{A.A.C.}},
\bauthor{\bsnm{{Sabhahit}}, \binits{G.N.}}:
\batitle{Maximum black hole mass across cosmic time}.
\bjtitle{Mon. Not. R. Astron. Soc.}
\bvolume{504}(\bissue{1}),
\bfpage{146}--\blpage{154}
(\byear{2021})
\end{barticle}
\endbibitem

\bibitem[\protect\citeauthoryear{{Renzo et al.}}{2019}]{2019A&A...624A..66R}
\begin{barticle}
\bauthor{\bsnm{{Renzo et al.}}}:
\batitle{Massive runaway and walkaway stars. A study of the kinematical imprints of the physical processes governing the evolution and explosion of their binary progenitors}.
\bjtitle{Astron. Astrophys.}
\bvolume{624},
\bfpage{66}
(\byear{2019})
\end{barticle}
\endbibitem


\bibitem[\protect\citeauthoryear{{Mapelli et al.}}{2020}]{2020ApJ...888...76M}
\begin{barticle}
\bauthor{\bsnm{{Mapelli et al.}}}:
\batitle{Impact of the Rotation and Compactness of Progenitors on the Mass of Black Holes}.
\bjtitle{Astrophys. J.}
\bvolume{888}(\bissue{2}),
\bfpage{76}
(\byear{2020})
\end{barticle}
\endbibitem


\bibitem[\protect\citeauthoryear{{Fern{\'a}ndez} et~al.}{2018}]{2018MNRAS.476.2366F}
\begin{barticle}
\bauthor{\bsnm{{Fern{\'a}ndez}}, \binits{R.}},
\bauthor{\bsnm{{Quataert}}, \binits{E.}},
\bauthor{\bsnm{{Kashiyama}}, \binits{K.}},
\bauthor{\bsnm{{Coughlin}}, \binits{E.R.}}:
\batitle{Mass ejection in failed supernovae: variation with stellar progenitor}.
\bjtitle{Mon. Not. R. Astron. Soc.}
\bvolume{476}(\bissue{2}),
\bfpage{2366}--\blpage{2383}
(\byear{2018})
\end{barticle}
\endbibitem

\bibitem[\protect\citeauthoryear{{Farag} et~al.}{2022}]{2022ApJ...937..112F}
\begin{barticle}
\bauthor{\bsnm{{Farag}}, \binits{E.}},
\bauthor{\bsnm{{Renzo}}, \binits{M.}},
\bauthor{\bsnm{{Farmer}}, \binits{R.}},
\bauthor{\bsnm{{Chidester}}, \binits{M.T.}},
\bauthor{\bsnm{{Timmes}}, \binits{F.X.}}:
\batitle{Resolving the Peak of the Black Hole Mass Spectrum}.
\bjtitle{Astrophys. J.}
\bvolume{937}(\bissue{2}),
\bfpage{112}
(\byear{2022})
\end{barticle}
\endbibitem

\bibitem[\protect\citeauthoryear{{Iorio et al.}}{2023}]{2023MNRAS.524..426I}
\begin{barticle}
\bauthor{\bsnm{{Iorio et al.}}}:
\batitle{Compact object mergers: exploring uncertainties from stellar and binary evolution with SEVN}.
\bjtitle{Mon. Not. R. Astron. Soc.}
\bvolume{524}(\bissue{1}),
\bfpage{426}--\blpage{470}
(\byear{2023})
\end{barticle}
\endbibitem

\bibitem[\protect\citeauthoryear{{Arca Sedda et al.}}{2023}]{2023MNRAS.526..429A}
\begin{barticle}
\bauthor{\bsnm{{Arca Sedda et al.}}}:
\batitle{The DRAGON-II simulations - II. Formation mechanisms, mass, and spin of intermediate-mass black holes in star clusters with up to 1 million stars}.
\bjtitle{Mon. Not. R. Astron. Soc.}
\bvolume{526}(\bissue{1}),
\bfpage{429}--\blpage{442}
(\byear{2023})
\end{barticle}
\endbibitem

\bibitem[\protect\citeauthoryear{{Costa et al.}}{2023}]{2023MNRAS.525.2891C}
\begin{barticle}
\bauthor{\bsnm{{Costa et al.}}}:
\batitle{Massive binary black holes from Population II and III stars}.
\bjtitle{Mon. Not. R. Astron. Soc.}
\bvolume{525}(\bissue{2}),
\bfpage{2891}--\blpage{2906}
(\byear{2023})
\end{barticle}
\endbibitem

\bibitem[\protect\citeauthoryear{{Santoliquido et al.}}{2020}]{2020ApJ...898..152S}
\begin{barticle}
\bauthor{\bsnm{{Santoliquido et al.}}}:
\batitle{The Cosmic Merger Rate Density Evolution of Compact Binaries Formed in Young Star Clusters and in Isolated Binaries}.
\bjtitle{Astrophys. J.}
\bvolume{898}(\bissue{2}),
\bfpage{152}
(\byear{2020})
\end{barticle}
\endbibitem

\bibitem[\protect\citeauthoryear{{Belczynski et al.}}{2020}]{2020A&A...636A.104B}
\begin{barticle}
\bauthor{\bsnm{{Belczynski et al.}}}:
\batitle{Evolutionary roads leading to low effective spins, high black hole masses, and O1/O2 rates for LIGO/Virgo binary black holes}.
\bjtitle{Astron. Astrophys.}
\bvolume{636},
\bfpage{104}
(\byear{2020})
\end{barticle}
\endbibitem

\bibitem[\protect\citeauthoryear{{Antonini} et~al.}{2018}]{2018MNRAS.480L..58A}
\begin{barticle}
\bauthor{\bsnm{{Antonini}}, \binits{F.}},
\bauthor{\bsnm{{Rodriguez}}, \binits{C.L.}},
\bauthor{\bsnm{{Petrovich}}, \binits{C.}},
\bauthor{\bsnm{{Fischer}}, \binits{C.L.}}:
\batitle{Precessional dynamics of black hole triples: binary mergers with near-zero effective spin}.
\bjtitle{Mon. Not. R. Astron. Soc. Lett.}
\bvolume{480}(\bissue{1}),
\bfpage{58}--\blpage{62}
(\byear{2018})
\end{barticle}
\endbibitem

\bibitem[\protect\citeauthoryear{{Stegmann} and {Klencki}}{2025}]{2025arXiv250609121S}
\begin{botherref}
\oauthor{\bsnm{{Stegmann}}, \binits{J.}},
\oauthor{\bsnm{{Klencki}}, \binits{J.}}:
{Spin-orbit misalignment and residual eccentricity are evidence that neutron star-black hole mergers form through triple star evolution}.
arXiv e-prints,
2506--09121
(2025)
\end{botherref}
\endbibitem


\bibitem[\protect\citeauthoryear{{De Luca} et~al.}{2025}]{2025arXiv250809965D}
\begin{botherref}
\oauthor{\bsnm{{De Luca}}, \binits{V.}},
\oauthor{\bsnm{{Franciolini}}, \binits{G.}},
\oauthor{\bsnm{{Riotto}}, \binits{A.}}:
{GW231123: a Possible Primordial Black Hole Origin}.
arXiv e-prints,
2508--09965
(2025)
\end{botherref}
\endbibitem

\bibitem[\protect\citeauthoryear{{Abac}}{2025}]{2025arXiv250708219T}
\begin{botherref}
\oauthor{\bsnm{{Abac}}, \binits{e.a.} \bsuffix{A.~G.}}:
{GW231123: a Binary Black Hole Merger with Total Mass 190-265 $M_{\odot}$}.
arXiv e-prints,
2507--08219
(2025)
\end{botherref}
\endbibitem

\bibitem[\protect\citeauthoryear{Rodriguez et~al.}{2016}]{Rodriguez2016c}
\begin{barticle}
\bauthor{\bsnm{Rodriguez}, \binits{C.L.}},
\bauthor{\bsnm{Zevin}, \binits{M.}},
\bauthor{\bsnm{Pankow}, \binits{C.}},
\bauthor{\bsnm{Kalogera}, \binits{V.}},
\bauthor{\bsnm{Rasio}, \binits{F.A.}}:
\batitle{Illuminating Black Hole Binary Formation Channels with Spins in Advanced LIGO}.
\bjtitle{Astrophys. J. Lett.}
\bvolume{832}(\bissue{1}),
\bfpage{L2}
(\byear{2016})
\end{barticle}
\endbibitem


\bibitem[\protect\citeauthoryear{{Mandel} et~al.}{2019}]{2019MNRAS.486.1086M}
\begin{barticle}
\bauthor{\bsnm{{Mandel}}, \binits{I.}},
\bauthor{\bsnm{{Farr}}, \binits{W.M.}},
\bauthor{\bsnm{{Gair}}, \binits{J.R.}}:
\batitle{Extracting distribution parameters from multiple uncertain observations with selection biases}.
\bjtitle{Mon. Not. R. Astron. Soc.}
\bvolume{486}(\bissue{1}),
\bfpage{1086}--\blpage{1093}
(\byear{2019})
\end{barticle}
\endbibitem


\bibitem[\protect\citeauthoryear{{Essick et al.}}{2025}]{44x3-hv3y}
\begin{barticle}
\bauthor{\bsnm{{Essick et al.}}}:
\batitle{Compact binary coalescence sensitivity estimates with injection campaigns during the LIGO-Virgo-KAGRA Collaborations' fourth observing run}.
\bjtitle{Phys. Rev. D}
\bvolume{112}(\bissue{10}),
\bfpage{102001}
(\byear{2025})
\end{barticle}
\endbibitem

\bibitem[\protect\citeauthoryear{{Essick} and {Farr}}{2022}]{2022arXiv220400461E}
\begin{botherref}
\oauthor{\bsnm{{Essick}}, \binits{R.}},
\oauthor{\bsnm{{Farr}}, \binits{W.}}:
{Precision Requirements for Monte Carlo Sums within Hierarchical Bayesian Inference}.
arXiv e-prints,
2204--00461
(2022)
\end{botherref}
\endbibitem

\end{thebibliography}


\newpage
\clearpage
\setcounter{section}{0}
\setcounter{figure}{0}

\renewcommand{\figurename}{Supplementary Figure}
\renewcommand{\tablename}{Supplementary Table}

\begin{center}
    {\LARGE\bfseries Supplementary Information\par}
\end{center}
\vspace{1cm}

\begin{figure*}[h!]
    \centering  
    \includegraphics[width=1.\textwidth]{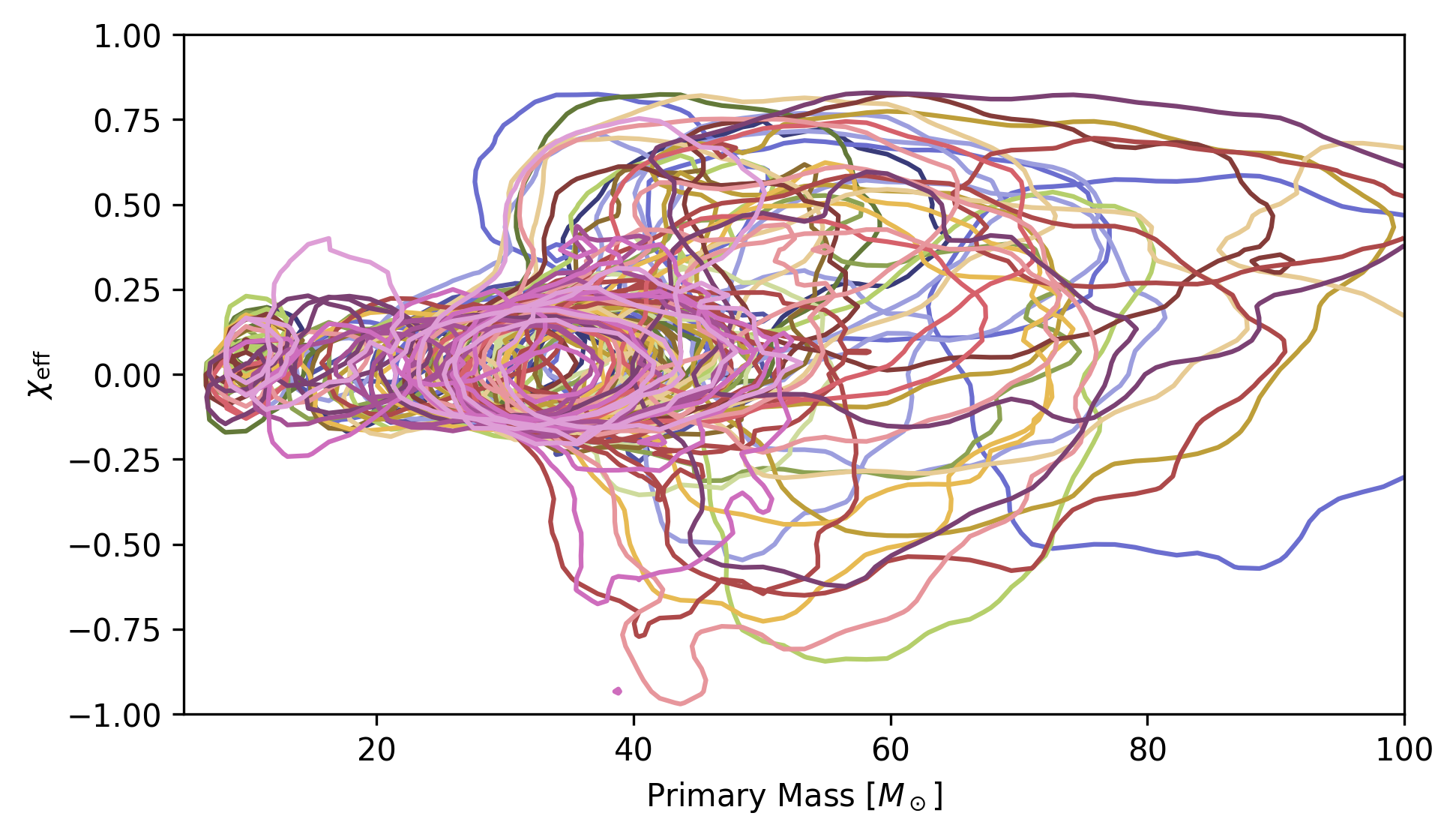}    
    \caption{Contours of the reweighted joint distribution of the primary black hole mass $m_1$ 
    and the effective inspiral spin parameter $\chi_\mathrm{eff}$. 
For each event, we compute a two-dimensional kernel density estimate over $(m_1, \chi_\mathrm{eff})$ from the reweighted posterior samples, and plot the 95\% credible region. 
Different colors correspond to different events.
 We fit the population
to a model where the  $\chi_{\rm eff}$  distribution is represented by a truncated Gaussian and a uniform distribution separated by mass.}
    \label{extfig2}
\end{figure*}
\begin{figure*}
	\centering
\includegraphics[width=0.452\textwidth]{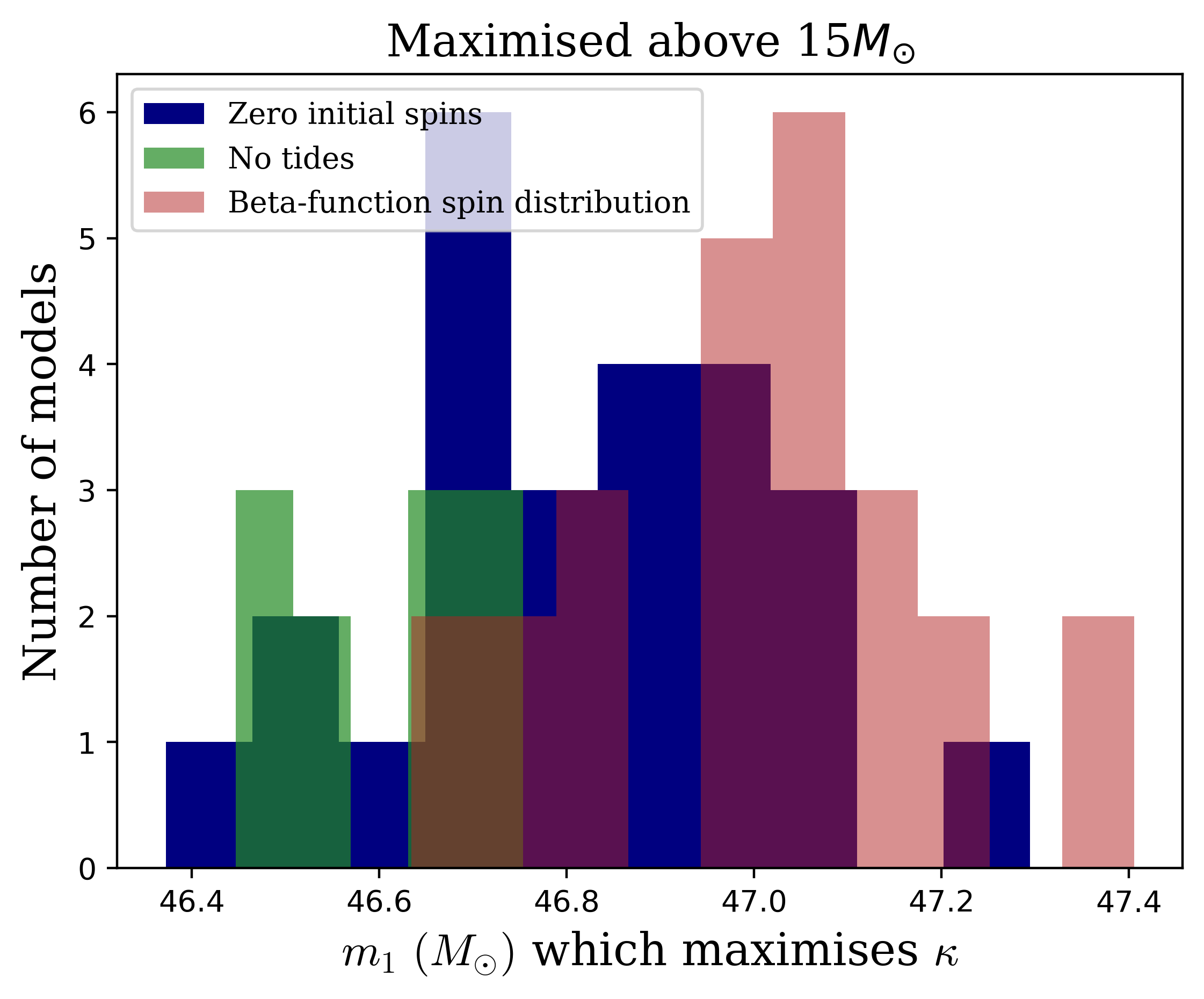}
	\includegraphics[width=0.45\textwidth]{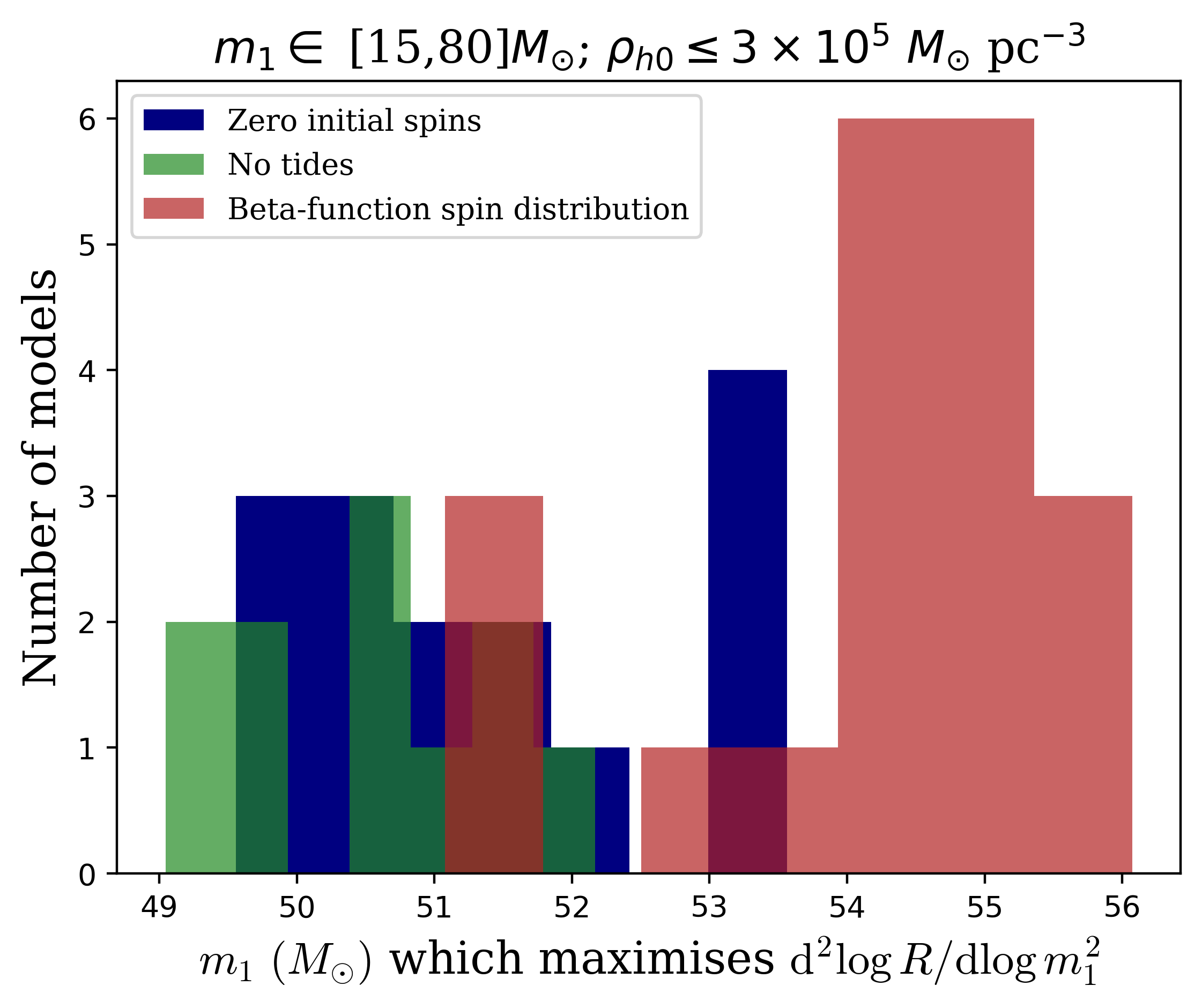}
	\caption{\emph{Left}: Histogram of the value of $m_1$ that maximizes $\kappa$ under various cluster population assumptions. 
\emph{Right}: Histogram of the value of $m_1$ that maximises $k$ for the same set of models, restricted to an 
initial half-mass density
$\rho_{h0} \leq 3 \times 10^{5}\, M_\odot\,\mathrm{pc}^{-3}$. 
Each point in the histogram corresponds to a realization of an evolved cluster population for  one of the 25 assumed initial half-mass densities.
For these models, the first-generation primary cutoff mass is fixed to $m_* = 50\,M_\odot$.
}
	\label{fig:second derivative histogram}
\end{figure*}

\begin{table}[h!]
\centering
 \renewcommand{\arraystretch}{1.05}
 \begin{tabular}{c l l}
 \hline
 \hline
 Parameter & Prior & Defined in \\
 \hline
  $a_{\chi}$
    & $\mathcal{HN}(3)$ & equation~4\\
    $\ln \ell_{\chi}$
    & $\mathcal{N}(-0.5,1)$ & equation~4 \\
 \hline
$a_{m}$ &$\mathcal{HN}(3)$ &  Mass model\\
  $\ln \ell_{m}$  & $\mathcal{N}(0,1)$ & Mass model  \\
  \hline
 $\tilde{m}$
    & $\mathcal{U}(20,100)$ & equations~5\\
    $\mu$  
    & $\mathcal{U}(-1,1)$ &  equations~4 and~8\\
    $\sigma$   
    & $\mathcal{LU}(-1.5,0)$ &  equations~4 and~8
     \\
     $\ln \ell_{\zeta}$
    & $\mathcal{N}(-0.5,1)$ & equation~6
    \\
   $a_{\zeta}$
    & $\mathcal{HN}(4)$ & equation~6
    \\
     $\beta_q$ & $\mathcal{N}(0,3)$ & equation~2 \\
$\kappa$ & $\mathcal{N}(0,6)$ & equation~3 \\
$\chi_{\rm eff,\;max}$ & $\mathcal{U}(0.05,1)$          & equation~8 \\
$\chi_{\rm eff,\;min}$ & $\mathcal{U}(-1,\chi_{\rm eff,\;max})$ & equation~8 \\
 \hline
 \hline
\end{tabular}
\caption{
Priors adopted for the hyperparameters of the population models. 
}
\label{tab:priors}
\end{table}
\clearpage

\newpage
\section{Uncertainties about the {pair instability} mass gap and alternative explanations for the spin transition }\label{PISN}
We interpret the spin transition and the cliff as connected with the physics of 
pair-instability supernovae  (PISN). In this Section, we discuss how this interpretation might be affected  by current uncertainties about stellar evolution and star cluster dynamics.

{  The theoretical origin of the PISN  is the rapid loss of radiation pressure that occurs once the oxygen-rich core of a massive star becomes hot enough for thermal photons to reach energies of order 
$\sim 1\,$MeV; at these energies, collisions between energetic photons and atomic nuclei can produce free electron–positron pairs, softening the equation of state. This triggers a thermal runaway in which the core contracts, heats further, and ignites oxygen explosively. Slightly less massive progenitors experience pulsational--PISN instead: repeated contractions and explosive oxygen-burning episodes eject substantial mass but do not fully unbind the star. The cumulative mass loss from these pulsations sets the lower edge of the PISN black-hole mass gap, as the resulting black holes are significantly lighter than their progenitor helium cores.}

{  The lower edge of the pair-instability mass gap is usually assumed to lie at
$40$--$50\,M_\odot$, based on calculations using pure-helium stellar models---i.e.,
stars that have lost their hydrogen envelopes from the zero-age main sequence
\cite{Woosley2016,2019ApJ...887...53F}. Although some early studies also examined
pair-instability in pure-oxygen stellar models \cite{1968Ap&SS...2...96F}, 
contemporary work typically adopts pure--helium
stellar models. The main motivation for this 
assumption is that massive stars are expected to lose their hydrogen envelopes either 
through binary interactions or, at sufficiently high metallicity, through strong 
line-driven winds.}
Moreover, codes integrating stellar structure encounter less numerical issues if the star does not develop a large hydrogen-rich envelope and/or a sharp core-envelope boundary \cite{2019ApJ...887...53F}.
However, several authors demonstrated that ---even for the fiducial ${}^{12}\mathrm{C}(\alpha,\gamma){}^{16}\mathrm{O}$ rate--- the lower edge of the mass gap can be located at a much higher value $70-90\,M_\odot$ in the case of a metal---poor {  single} star that retains a large portion of its H-rich envelope until collapse \cite{Woosley2016,2019ApJ...887...53F,Woosley2021,2023MNRAS.526.4130H}. Here, {  low metallicity ($Z<10^{-3}$) and isolated evolution } imply suppressed stellar winds and hence survival of the envelope \citep{2021MNRAS.504..146V}. Other uncertainties stem from assumptions for convection \cite{2019A&A...624A..66R}, core overshooting and envelope undershooting that can lead to substantial dredge-up episodes
\cite{2021MNRAS.501.4514C}, stellar rotation \cite{2020ApJ...888...76M}, mass loss during pulsational pair instability \cite{2019ApJ...878...49W}, and onset of shocks during failed supernovae \cite{2018MNRAS.476.2366F}. 

{ 
Scenarios that place the lower edge of the gap significantly above 
$\sim 60\,M_\odot$ typically require the progenitor to retain at least part of its 
hydrogen envelope until collapse (but see \cite{2022ApJ...937..112F} for a counterexample). This outcome is unlikely in tight binary systems, 
where envelope stripping during mass transfer naturally produces a sharp drop in the 
primary-mass distribution above $\sim 40\,M_\odot$ 
\cite[e.g., Fig.~12 in][]{2023MNRAS.524..426I}. Thus, we do not expect isolated 
binary evolution to populate the region above the cliff. In contrast, very massive 
metal-poor single stars or stellar-collision products 
\cite[e.g.,][]{2023MNRAS.526..429A} can form black holes with 
masses $\gtrsim 50\,M_\odot$ \cite[e.g.,][]{2023MNRAS.525.2891C}, and these channels 
could also {}populate the valley near $\simeq 14\,M_\odot$.
}

Dynamical processes (especially exchanges) in a dense stellar cluster are the most effective mechanisms through which such over-sized single black holes might pair up with other black holes and merge. This is also compatible with a cliff in the merger rate density at $40\,M_\odot$, as dynamical exchanges involving over-sized black holes 
 are orders of magnitude less common than isolated binary mergers, even in optimistic cases \cite{2020ApJ...898..152S}.

The most controversial aspect is the black hole spin distribution of such oversized black holes. While there are several reasons to suspect that high spins are possible in this scenario (e.g., limited ejection of mass and angular momentum at low metallicity, spin up of a stellar collision product), a study that investigates the spins of oversized first generation black holes is missing. Considering such uncertainties, we will extend our analysis to first-generation oversized black holes in future works.

{  Finally, we note  that binary mass transfer can produce a cutoff in the secondary-mass distribution at $\sim 40\,M_\odot$ without PISN mass gap \cite[e.g.,][]{2020A&A...636A.104B}. Triple interactions could also lead to misaligned systems \cite[e.g.,][]{2018MNRAS.480L..58A,2025arXiv250609121S}, 
and high spins above the mass gap through consecutive mergers \cite{2021MNRAS.502.2049L}, although these latter are thought to be extremely rare. }
Finally,  primordial black holes are another viable interpretation for the population above the cliff. Although traditionally associated with low spins, recent models show that primordial black holes might also achieve large spins
\cite{2025arXiv250809965D}.

\section{$\chi_{\rm eff}$ distribution as a Gaussian process}

\begin{figure*}
    \centering  
\includegraphics[width=1.\textwidth]{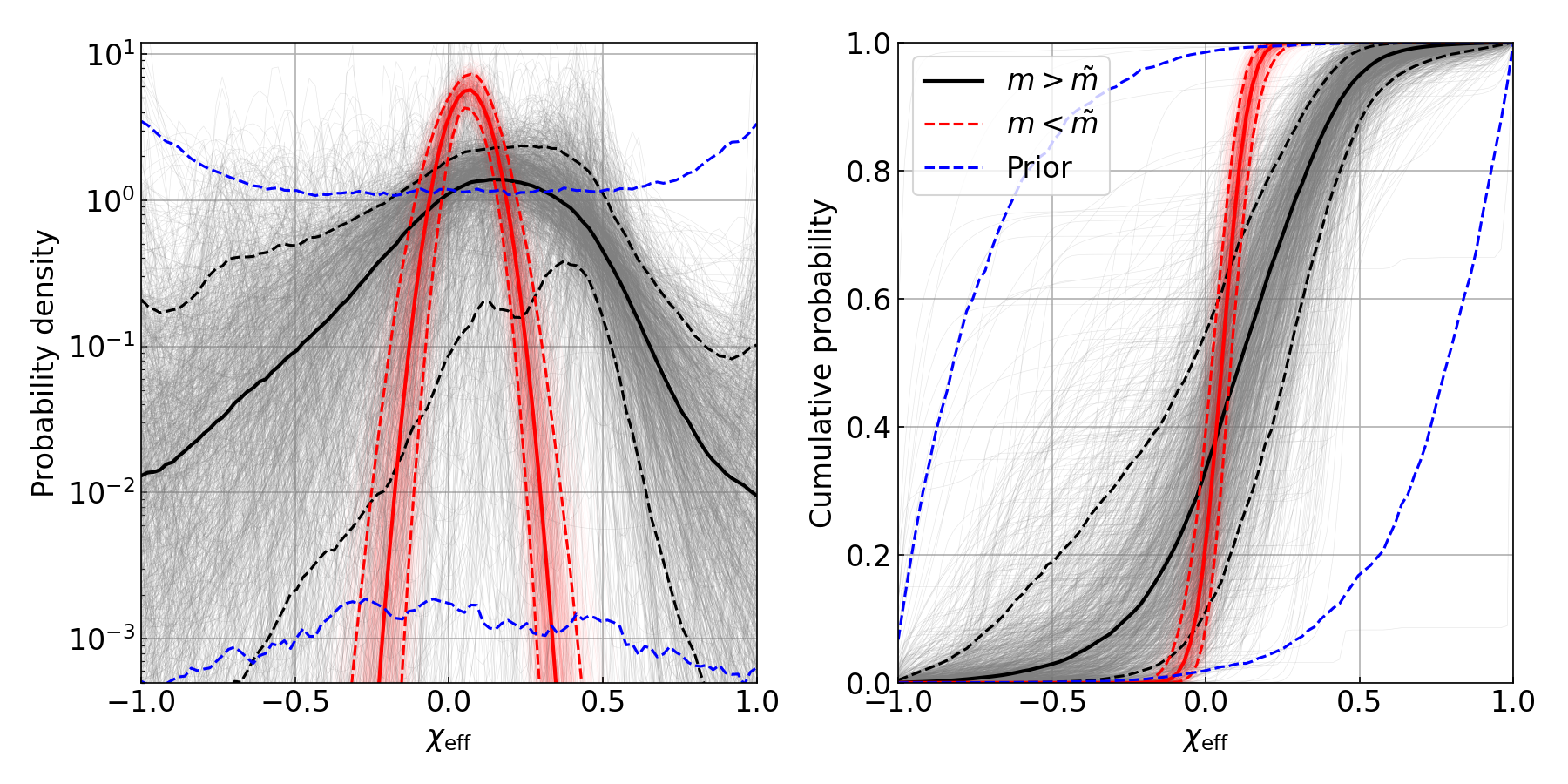}
    \caption{
The distribution of $\chi_{\rm eff}$ for the low and high mass populations under equation~4 in the Methods section. Here the $\chi_{\rm eff}$ distribution of the high-mass  population is modeled non-parametrically.
Solid lines are median, while dashed lines show 10\% and 90\% of the distributions.
The population below $\tilde{m}$ is represented by a narrow Gaussian inconsistent with  hierarchical formation, while the population above  $\tilde{m}$ is characterized by a broad
$\chi_{\rm eff}$ distribution that is  consistent with  isotropy (i.e., symmetry around zero) as expected for hierarchical mergers in dynamical environments \cite{Rodriguez2016c}.
}
    \label{non-par}
\end{figure*}

Supplemetary Figure~\ref{non-par} shows the inferred distribution of $\chi_{\rm eff}$ under equation~4 in the Methods section where the $\chi_{\rm eff}$
distribution of the high-mass population is treated non-parametrically using a Gaussian Process prior.
For $m_1 < \tilde{m}$, the distribution is well described by a narrow Gaussian with mean $\mu = 0.04^{+0.02}_{-0.02}$ and standard deviation $\log_{10} \sigma=-1.15^{+13}_{-15}$, while for $m_1 > \tilde{m}$ it broadens and becomes consistent with being a uniform distribution that is symmetric around zero.
We estimate the posterior probability that the effective spin distribution extends to negative values by evaluating CDF($\chi_{\mathrm{eff}}=0$) for each posterior sample of the non-parametric model. Because the model probability is non-zero across $\chi_{\mathrm{eff}}$, we compute  the fraction of samples with 
CDF$( \chi_{\mathrm{eff}}=0)>0.01$. This provides a conservative proxy for testing whether the minimum supported $\chi_{\mathrm{eff}}$ value is negative. We find that more than $98\%$ of posterior samples satisfy this condition, indicating that the non-parametric inference strongly favors a high-mass binary population that includes systems with negative effective spin.

This is a consistent result recovered among the models presented in this work and shows that a distribution which only contains binaries with aligned spins for the high-mass population is  disfavored by current data.
We also obtain ${\rm CDF}(\chi_{\rm eff}=0) =
0.29^{+0.30}_{-0.26}$ ($90\%$ credibility),  indicating that current observations do not  place stringent constraints on the symmetry of the
$\chi_{\rm eff}$ distribution based on this model. 

{  Finally, we note that excluding the exceptionally massive event GW231123 \cite{2025arXiv250708219T} from our sample, which may involve other channels other than hierarchical mergers or which primary might lie above the PISN mass gap, leads to  tighter constraints on the presence of misaligned systems in the high-mass  population. In this case, we find that the population contains misaligned systems with negative $\chi_{\rm eff}$ at $99\%$ confidence and that ${\rm CDF}(\chi_{\rm eff}=0) = 0.31^{+0.28}_{-0.27}$.}



\section{Hierarchical inference}\label{Hinf}

We carry out hierarchical inference on the binary black hole population using Hamiltonian Monte Carlo (HMC) in \texttt{numpyro}, a probabilistic programming framework built on \texttt{jax}. 
Within the standard framework of hierarchical Bayesian inference, for each event with posterior $p(\theta_i|d_i)$, the hyperparameter posterior is \cite[e.g.,][]{Fishbach_2018,2019MNRAS.486.1086M,2022ApJ...937L..13C}
\begin{equation}
    p(\Lambda|\{d_i\}) \propto p(\Lambda)\,\xi^{-N_{\rm obs}}(\Lambda)
    \prod_{i=1}^{N_{\rm obs}}
    \Big\langle \frac{p(\theta_i|\Lambda)}{p_\mathrm{pe}(\theta_i)} \Big\rangle,
    \label{eq:likelihood-sum-short}\nonumber
\end{equation}
where $p_\mathrm{pe}(\theta_i)$ is the prior used in parameter estimation and $\langle\cdot\rangle$ denotes an expectation over posterior samples.

The detection efficiency $\xi(\Lambda)$ is computed using injection campaigns \cite{2022ApJ...937L..13C,2024PhRvD.109b2001A,injections,44x3-hv3y}, 
\begin{equation}\nonumber
    \xi(\Lambda) = \frac{1}{N_{\rm inj}} \sum_{i=1}^{N_{\rm found}} 
    \frac{p(\theta_i|\Lambda)}{p_\mathrm{inj}(\theta_i)}\, ,
    \label{eq:csi-short}
\end{equation}
where injections are reweighted from the reference distribution $p_\mathrm{inj}$ to the proposed model $p(\theta|\Lambda)$.

To mitigate sampling variance, we track the effective number of posterior samples \cite{2022arXiv220400461E},
\begin{equation}\nonumber
    N_{\mathrm{eff},i}(\Lambda) = 
    \frac{\left[\sum_j w_{i,j}(\Lambda)\right]^2}{\sum_j w_{i,j}^2(\Lambda)}\ ,
    \label{eq:Neff-short}
\end{equation}
with $w_{i,j}(\Lambda) = p(\theta_{i,j}|\Lambda)/p_\mathrm{pe}(\theta_{i,j})$, and the effective number of injections,
\begin{equation}\nonumber
    N_\mathrm{eff}^\mathrm{inj}(\Lambda) = 
    \frac{\left(\sum_i w_i(\Lambda)\right)^2}{\sum_j w_j^2(\Lambda)}\, .
\end{equation}
Following \cite{2022arXiv220400461E}, we require $N_\mathrm{eff}^\mathrm{inj} \gtrsim 4N_\mathrm{obs}$. 
We safeguard the inference by penalizing models with $N_\mathrm{eff}^\mathrm{inj} < 4N_\mathrm{obs}$ or 
$\min \log N_{\mathrm{eff},i} < 0.6$, adding
\begin{equation}\nonumber
    \ln S\!\left(\tfrac{N_\mathrm{eff}^\mathrm{inj}}{4N_\mathrm{obs}}\right) +
    \ln S\!\left(\tfrac{\mathcal{N}}{0.6}\right), \qquad
    S(x)=\tfrac{1}{1+x^{-30}},
\end{equation}
to the log-likelihood. This ensures that models with pathologically low effective sample sizes are excluded.

\end{document}